\documentclass[pra, twocolumn,showpacs,nofootinbibfloatfix,amsmath,amsfonts,amssymb]{revtex4-1}%
\usepackage{amsmath,amsfonts,amssymb,color}
\usepackage{amsthm}
\usepackage{leftidx}
\usepackage{graphicx}
\usepackage{xcolor}
\usepackage{dcolumn}
\usepackage{bm}
\usepackage{epstopdf}
\usepackage{epsfig}
\usepackage{mathdots} 
\def\ch{\textcolor{black}}

\begin{document}
	\title{Exploring topological phases with extended Su-Schrieffer-Heeger models}
	\author{Raditya Weda Bomantara}
	\email{Raditya.Bomantara@kfupm.edu.sa}
	\affiliation{%
		Department of Physics, Interdisciplinary Research Center for Advanced Quantum Computing, King Fahd University of Petroleum and Minerals, 31261 Dhahran, Saudi Arabia
	}
	\date{\today}
	
	%%%%%%%%%%%%%%%%%%%% ABSTRACT %%%%%%%%%%%%%%%%%%%%%%%%
	%\begin{linenumbers}
	
	\vspace{2cm}
	
\begin{abstract}
The Su-Schrieffer-Heeger (SSH) model describes a tight-binding one-dimensional (1D) lattice with alternating nearest-neighbor amplitudes. Despite its mathematically simple and physically intuitive structure, the SSH model is capable of supporting a 1D topological phase that is characterized by the presence of zero energy eigenstates (zero modes) localized at each end of the lattice. For this reason, many studies in the area of topological phases of matter often consider the SSH model as a subject for various extensions that give rise to more sophisticated topological phenomena. The purpose of this article is to review, in sufficient detail, existing approaches to extending the SSH model. This includes extensions by increasing the dimensionality of the lattice, enlarging the size of its unit cell, or adding extra terms that represent various physical effects. For each approach, some extended SSH models studied in relevant existing literature are discussed as case studies. Noteworthy properties of such models, which are of topological origin, are further comprehensively elaborated.   
	
\end{abstract}

\maketitle

\section{Introduction} 
\label{intro}

Topological phases of matter are exotic systems exhibiting physical properties that are highly resistant to local perturbations. Such systems first gained popularity in the 2000s following seminal theoretical \cite{Jungwirth02,Onoda03,Liu08,Yu10,Kane05,Bernevig06,Kane05b} and experimental \cite{konig07,Hsieh08,Zhang09,Xia09,Chen09} studies. The interest in topological phases of matter is further amplified upon realizing their potential applications for quantum computations \cite{Kitaev03,Nayak08}. Since then, topological phases of matter have been actively studied from different angles, some focusing on the more fundamental aspects such as their mathematical characterization based on the underlying symmetries \cite{Schnyder08,Ryu10,Roy17} or their implementation in various physical platforms \cite{Wang18,Chen18,Xiao15,Stuhl15,Meier16,Goldman16,Jotzu14,Meier18,Li18,Kiczynski22,Ocampo24}, whilst others are on the more applied sides such as their application in quantum transport \cite{Kraus12,Verbin15,Longhi19,Lohse16,Nakajima16,Wang13,Mei14,Wang15,Ke16,Yap17,Ke17,Hu19,Lin20,Hu20} or their use as quantum technological devices \cite{Lang17,Karzig17,Mei18,Bomantara18,Boross19,Tan20,Bomantara20,Bomantara20b,Matthies22,Safwan25}.

Topologically protected edge modes, which refer to a class of eigenstates that are localized near the system's boundaries, are the typical defining observables of a topological phase. Depending on the dimensionality and symmetries of the system at hand, these edge modes could be either chiral or nonchiral in nature. Chiral edge modes are associated with linearly dispersing energy in a two-dimensional (2D) \cite{Sun11,Weeks10,Guo09,Lindner11,Ho14} or a three-dimensional (3D) \cite{Fu07,Moore07,Roy09,Roushan09,Hasan10,Ringel12} topological system. By contrast, nonchiral edge modes are pinned at a particular energy due to some protecting symmetries \cite{Kitaev01,Fendley12,Ganeshan13,Bomantara16,Bomantara16b,Wang17}. In all these cases, there exist appropriate topological invariants, the values of which correlate with the corresponding number of edge modes \cite{Prodan16}, thereby highlighting their topological nature.

The Su-Schrieffer-Heeger (SSH) model \cite{Su79} describes a minimal one-dimensional (1D) lattice configuration that supports a topological phase. Specifically, the SSH Hamiltonian consists of nearest-neighbor hopping terms with alternating amplitudes. Depending on the relative strength between the two hopping amplitudes, zero energy edge modes may emerge at each end of the lattice \cite{Su79}. When the SSH model is defined on a ring geometry, i.e., the two ends of the lattice are identified, the Zak phase \cite{Zak88} associated with the system's eigenstates can be shown to be proportional to a winding number invariant \cite{Delplace11}, which takes a quantized value of either zero or unity. Interestingly, the condition on the two hopping amplitudes which yields a winding number of zero (unity) coincides with that which results in the absence (presence) of zero energy edge modes. This correlation not only highlights the topological origin of zero energy edge modes in the SSH model, but also confirms the theory of bulk--boundary correspondence \cite{Jackiw75,Wen94}.   

Due to its simplicity, the SSH model often forms a starting point for developing more sophisticated topological models. For example, the Qi-Wu-Zhang (QWZ) model \cite{Qi06} could be viewed as an extension of the SSH model to a 2D lattice which supports a pair of chiral edge modes in its topologically nontrivial regime. Further extension of the model to a 3D lattice will then result in either a 3D topological insulator \cite{Shen17} or the so-called Weyl semimetal \cite{Yang11}. Recently, the concept of higher-order topological phases is established \cite{Benalcazar17,Benalcazar17b,Song17,Langbehn17,Schindler18}. Specifically, a $d$th-order topological phase refers to a type of topological phase that supports topological modes in the $(N-d)$-boundaries of the system, where $N$ is the dimension of the system. Intriguingly, the simplest forms of higher-order topological phases can be obtained by appropriately stacking SSH chains to form either a 2D or 3D lattice \cite{Li18b,Bomantara19}.

Extension of the SSH model can be made even without changing the dimensionality of the lattice. This could be accomplished, e.g., by changing the periodicity of the hopping amplitude. Indeed, a number of studies \cite{Anas22,Alv19,Du24,Zhou23,Marq20} have proposed and extensively studied a class of extended SSH models whose hopping amplitudes repeat every $n>2$ sites. By construction, such models exhibit a larger number of energy bands and different protecting symmetries as compared with the regular SSH model, which consequently result in richer edge mode structure. This enlargement in the number of energy bands can also be obtained through some other means, e.g., by applying a square-rooting procedure \cite{Ark17,Ezawa20,Marq21,Bomantara22,Zhou22} reminiscent of the transformation from the Klein-Gordon equation to the Dirac equation \cite{Dirac28} or by replacing the (pseudo)-spin degrees of freedom with their higher-level counterparts \cite{Ghun24,Ghun25,Ghun25b}.

Another approach for extending the SSH model involves adding extra terms to the SSH Hamiltonian that represent various physical effects such as periodic driving \cite{Asb14,Lago15,Zhou18}, non-Hermiticity \cite{Yao18,Lieu18,Ye24}, long range hopping \cite{Hsu20,Gonz19,Mal23,Cha25}, few- or many-body interactions \cite{Liberto16,Salvo24}, and/or nonlinearity \cite{Tul20,Ezawa21}. It is worth noting that the interplay between topology and each of these effects is often subject to an active research area of its own. For example, a periodically driven SSH model yields a type of Floquet topological phase \cite{Kitagawa10,Cay13,Po16,Pot17,Umer20,Umer21} that exhibits the so called $\pi$ modes \cite{Jiang11,Bomantara18b,Bomantara20c,Zhu22}, i.e., topological edge modes that are pinned at half the driving frequency, which have no static counterparts. The other effects similarly enrich the edge mode profiles of the resulting extended SSH models \cite{Zhou17,Ma21,Parto18,Gun22,Ghosh25,Marques17}.

In this review article, the various extensions of the SSH model above will be elaborated in detail. While the discussion of such extended models often touch on different timely research areas, it should be emphasized that our focus is on highlighting the topological origin of each extended SSH model by drawing analogy with the regular SSH model. The main focus of this article is then to highlight the potential of the seemingly simple SSH model to generate and systematically study more complex topological phases.

The remainder of this article is structured as follows. In Sec.~\ref{overview}, the SSH model is reviewed in sufficient detail. To this end, its explicit form of the second-quantized Hamiltonian will first be presented, along with its intuitive interpretation. Its nontrivial topology is then uncovered by computing the appropriate topological invariant and explicitly showing the subsequent emergence of topological edge modes. In Sec.~\ref{TISSH}, the topological structure of the SSH model is linked to that of a modified 1D Dirac equation in the continuum limit. By considering the higher-dimensional versions of the modified Dirac equation and discretizing the corresponding Hamiltonians, a number of higher-dimensional topological systems are obtained. Sec.~\ref{HOSSH} presents a different higher-dimensional extension of the SSH model, which amounts to stacking multiple copies of the SSH Hamiltonians, to yield a class of higher-order topological phases. In Sec.~\ref{SSH3}, extended SSH models that arise from enlarging the sublattice size while keeping the structure of the real space Hamiltonian intact is discussed. In Sec.~\ref{SSH3M}, a different unit cell enlargement of the SSH model, obtained by replacing the Pauli matrices in the momentum space Hamiltonian of the SSH model by their higher-level counterparts, is elaborated in detail. Sec.~\ref{SSHeff} elucidates some extended SSH models and their notable topological properties that arise from incorporating periodic driving, non-Hermiticity, longer-range hopping, interaction effect, or nonlinearity. Finally, we conclude this review and discuss future prospects of utilizing the SSH model and its extensions for studying exotic topological phases in Sec.~\ref{conc}.  

\section{Overview of the SSH model}
\label{overview}

\begin{center}
    \begin{figure}
        \centering
        \includegraphics[scale=0.5]{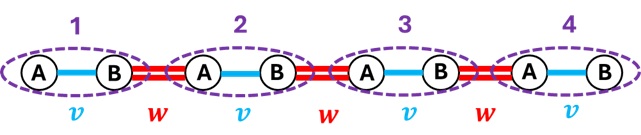}
        \caption{Schematic description of the SSH Hamiltonian with $N=4$. Each unit cell (dashed ellipse) consists of two sublattices labeled as A and B.}
        \label{fig:sshdep}
    \end{figure}
\end{center}

The SSH model was first introduced in Ref.~\cite{Su79} as a simplistic description of a polyacetylene polymer. In the second quantized notation, the corresponding Hamiltonian takes the form
\begin{equation}
    H = \sum_{j=1}^{N} v c_{j,B}^\dagger c_{j,A} +\sum_{j=1}^{N-1} w c_{j+1,A}^\dagger c_{j,B} +h.c. , \label{eq:SSH}
\end{equation}
where $v$ and $w$ are real parameters that respectively represent the intra- and intercell hopping parameters, $N$ is the length of the chain, and $c_{j,S}$ ($c_{j,S}^\dagger$) describes the particle annihilation (creation) operator at sublattice $S=A,B$ of site $j$. Figure~\ref{fig:sshdep} depicts the schematic description of the SSH Hamiltonian, i.e., Eq.~(\ref{eq:SSH}).

\begin{center}
    \begin{figure}
        \centering
        \includegraphics[scale=0.4]{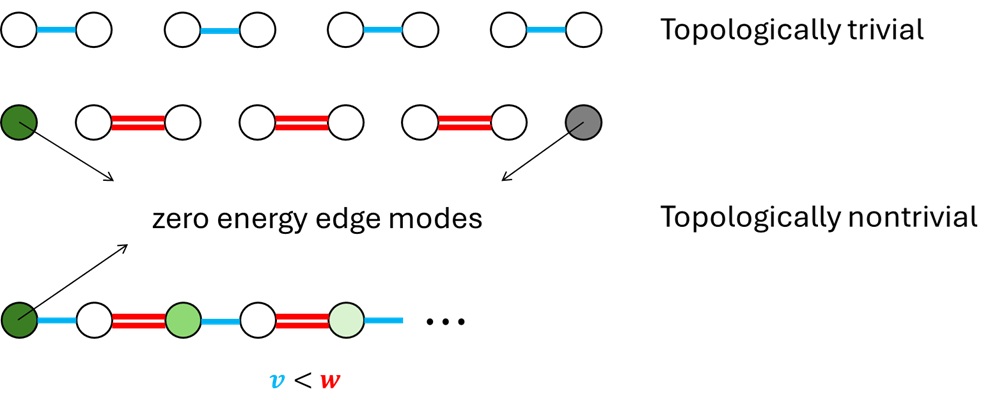}
        \caption{Schematic description of the SSH Hamiltonian in the topologically trivial and nontrivial regime, as well as the zero energy edge modes in the latter.}
        \label{fig:sshdep2}
    \end{figure}
\end{center}

The SSH Hamiltonian is known to support two distinct topological phases, characterized by the presence and absence of zero energy edge modes. Intuitive insight into these phases can be obtained by considering two limiting cases. First, at $v=0$, it is easily checked that two operators $c_{1,A}$ and $c_{N,B}$ commute exactly with the SSH Hamiltonian. Moreover, since $c_{1,A}$ and $c_{N,B}$ respectively occupy the left- and rightmost site of the lattice, these operators are termed the zero energy edge modes. By contrast, at $w=0$, operators with the same properties do not exist, thereby implying the absence of the zero energy edge modes. These two cases are schematically depicted in the upper and middle panels of Fig.~\ref{fig:sshdep2}.

While not immediately obvious visually, zero energy edge modes remain present as long as the hopping amplitudes obey the condition $v<w$ and the lattice is infinitely long. Indeed, it is easily verified that the linear combination 
\begin{equation}
    c_{0}=\sum_{j=1}^\infty \left(-\frac{v}{w}\right)^j c_{j,A} ,
\end{equation}
commutes exactly with the Hamiltonian. Since $v<w$ was assumed, the coefficient $\left(-\frac{v}{w}\right)^j \rightarrow 0$ as $j\rightarrow \infty$, which implies that $c_0$ is localized near the left end of the lattice (see the bottom panel of Fig.~\ref{fig:sshdep2}). That is, $c_0$ is a zero energy edge mode. 

In a finite lattice of length $N$, the series in $c_0$ necessarily terminates. Consequently, the commutator between $c_0$ and the Hamiltonian is not exactly zero, but it yields a term that decays exponentially with $N$. Strictly speaking, unless $v=0$, $c_0$ only represents an approximate zero energy edge mode of the system, the localization length of which depends on the ratio $\frac{v}{w}$. Physically, the absence of an exact zero energy edge mode in a finite lattice is attributed to the hybridization, i.e., ``coupling", between the edge modes localized at the opposite edges due to their overlap in the bulk. Indeed, a second approximate zero energy edge mode that is localized near the right end of the lattice can be similarly constructed via the linear combination
\begin{equation}
    c_{0}'=\sum_{j=1}^N \left(-\frac{v}{w}\right)^{N-j} c_{j,B} .
\end{equation}

The topological origin of the above zero energy edge modes can further be uncovered by studying the SSH Hamiltonian under periodic boundary conditions (PBC), where $c_{1,S}=c_{N+1,S}$. To this end, we first define \ch{
\begin{equation}
    c_{k,S} = \frac{1}{\sqrt{N}} \sum_{j=1}^N c_{j,S} e^{\mathrm{i} k j}, \label{eq:rtmtrans}
\end{equation}
}where $k=-\pi+\frac{2\pi m}{N}$ for $m=1,2,\cdots,N$ \ch{and $S=A,B$}. The SSH Hamiltonian can then be written as
\begin{equation}
    H=\sum_{k} \psi_k^\dagger \mathcal{H}(k) \psi_k ,
\end{equation}
where $\psi_k =\left(c_{k,A},c_{k,B}\right)^T$, 
\begin{equation}
    \mathcal{H}(k) \equiv \left(v+w \cos(k)\right) \sigma_x + w \sin(k) \sigma_y , \label{eq:SSHm}
\end{equation}
is the corresponding momentum space Hamiltonian, $\sigma_x$ and $\sigma_y$ are the Pauli matrices acting on the sublattice subspace. In the limit $N\rightarrow \infty$, $k$ takes a continuous value within $(-\pi,\pi]$, and one may define the Zak phase \cite{Zak88} as
\begin{equation}
    \gamma_\pm = \mathrm{i} \int_{-\pi}^\pi \langle \phi_\pm^\dagger(k) |\frac{\partial}{\partial k} |\phi_\pm(k) \rangle dk , \label{Eq:zak}
\end{equation}
where \ch{$|\phi_\pm(k) \rangle =\frac{1}{\sqrt{2}}\left(1,\pm e^{\mathrm{i} \xi}\right)^T$} with \ch{$\xi=\arctan\left(\frac{w \sin(k)}{v+w\cos(k)}\right)$} is the eigenstate of $\mathcal{H}(k)$ associated with $E_\pm (k) = \pm \sqrt{v^2 +w^2 +2vw \cos(k)}$. By replacing $ w e^{\mathrm{i} k} \rightarrow z$, we can turn $\gamma_\pm$ into the complex integration
\begin{eqnarray}
    \gamma_{\pm} &=& \pi \mathcal{W} , \nonumber \\
    \mathcal{W} &=& \frac{1}{2\pi \mathrm{i}} \oint_\mathcal{C} \frac{1}{v+z} dz , \label{eq:ZakSSH}
\end{eqnarray}
where $\mathcal{C}$ is a (counterclockwise) circular path of radius $w$ in the complex plane, centered around the origin, and $\mathcal{W}$ is mathematically known as the winding number. By construction, the winding number can only take an integer value and usually serves as a topological invariant. Indeed, by applying Cauchy's residue theorem, it is easily verified that \ch{$\mathcal{W} = 1$ whenever $v<w$ and $\mathcal{W} = 0$ otherwise}. Consequently,
\begin{equation}
    \gamma_\pm =\begin{cases}
        0 & \text{ for }v>w \\
        \pi & \text{ for }w>v
    \end{cases} .
\end{equation}
Remarkably, the two values of Zak phase above perfectly correlate with the presence/absence of zero energy edge modes as previously obtained from directly analyzing the system under open boundary conditions (OBC). That is, $\gamma_\pm = 0$ ($\gamma_\pm = \pi$) corresponds to a topologically trivial (nontrivial) phase that is marked by the absence (presence) of zero energy modes.

The winding number $\mathcal{W}$ of Eq.~(\ref{eq:ZakSSH}) can also be obtained from the momentum space Hamiltonian directly by exploiting its symmetries. To this end, it is first worth noting that $\mathcal{H}$ possesses time-reversal, particle-hole, and chiral symmetries that satisfy, respectively,
\begin{eqnarray}
    \mathcal{T}^{-1} \mathcal{H}(k) \mathcal{T} &=& \mathcal{H}(-k) , \nonumber \\
    \mathcal{P}^{-1} \mathcal{H}(k) \mathcal{P} &=& -\mathcal{H}(-k) , \nonumber \\
    \mathcal{C}^{-1} \mathcal{H}(k) \mathcal{C} &=& -\mathcal{H}(k),
\end{eqnarray}
where $\mathcal{T}=\mathcal{K}$ ($\mathcal{K}$ being the complex conjugate), $\mathcal{P}=\mathcal{K}\sigma_z$, and $\mathcal{C}=\sigma_z$. In the Altland-Zirnbauer (AZ) symmetry classification \cite{Altland97}, the SSH model belongs to the BDI class, which is characterized by an integer topological invariant. Specifically, these symmetries manifest themselves in the block anti-diagonal structure of the momentum space Hamiltonian, i.e., 
\begin{equation}
    \mathcal{H}(k) = \left( \begin{array}{cc}
       0  & h_-(k) \\
       h_+(k)  & 0 
    \end{array} \right),
\end{equation}
where \ch{$h_\pm (k) = v + w e^{\pm \mathrm{i} k}$}. In this case, the winding number of the anti-diagonal element $h_+$ is precisely $\mathcal{W}$ of Eq.~(\ref{eq:ZakSSH}).

\section{Extending the SSH model by going to a higher dimension}

\subsection{Understanding higher-dimensional topological phases from the SSH model}
\label{TISSH}

\ch{We start by considering a modified 1D Dirac Hamiltonian with a Wilson-type (quadratic-in-momentum) mass term \cite{Shen17}, i.e., }
\begin{equation}
    \mathcal{H}_{\rm cont}(p) \equiv (A-B p^2) \sigma_x + C p \sigma_y ,
\end{equation}
where $A$, $B$, and $C$ are some real constants. In particular, note that the eigenstates of the $\mathcal{H}_{\rm cont}(p)$ at $p=0$ are the same as those of $\sigma_x$, i.e., $|\pm\rangle$ associated with \ch{eigenvalues of} $\pm 1$. If $AB<0$, a state initially prepared in $|+\rangle$ at $p=0$ ends up in the same state at $p\rightarrow \infty$. Meanwhile, if $AB>0$, a state initially prepared in $|+\rangle$ at $p=0$ ends up in the other eigenstate $|-\rangle$ at $p\rightarrow \infty$. Therefore, the two eigenstates' structures as $p$ smoothly changes from $0$ to $\infty$ are topologically distinct. The case $AB<0$ is classified as topologically trivial due to the structural equivalence of the system's eigenstates with those of vacuum or constant mass Dirac Hamiltonian, while the case $AB>0$ corresponds to the topological nontrivial regime. As an immediate consequence of this classification, a finite system described by $\mathcal{H}_{\rm cont}(p)$ with $AB>0$ supports boundary states at the interface with vacuum or other topologically trivial systems \cite{Shen17}, while that with $AB<0$ does not.    

\ch{It follows that the SSH model represents the tight-binding discretization of $\mathcal{H}_{\rm cont}(p)$ to define it on a 1D lattice structure. Specifically, this amounts to replacing 
\begin{equation}
p\equiv -\mathrm{i}\frac{\partial}{\partial x}\rightarrow \mathrm{i}\sum_{j=1}^{N-1} \left(|j\rangle \langle j+1|-|j+1\rangle \langle j|\right) ,    
\end{equation}
and
\begin{equation}
    p^2 \equiv -\frac{\partial^2}{\partial x^2} \rightarrow  \left[\sum_{j=1}^{N-1}\left(|j\rangle \langle j+1|+|j+1\rangle \langle j|\right) -2 \sum_{j=1}^N |j\rangle \langle j|\right]
\end{equation}
where $|j\rangle$ represents the basis (first-quantized) state associated with site $j$. By further interpreting $\sigma_x$ and $\sigma_y$ as acting on the sublattice degree of freedom, as well as taking $A-2B=v$ and $B=C=w/2$, the first-quantized version of the SSH model is obtained. By applying the above procedure of discretization on some higher-dimensional variants of the continuum Hamiltonian, a variety of higher-dimensional topological phases that serve as extensions of the SSH model can be obtained.  
}

\subsubsection{Extension to 2D topological insulators}

Generalizing $\mathcal{H}_{\rm cont}(p)$ to 2D is straightforward, i.e.,
\begin{eqnarray}
    \mathcal{H}^{(2D)}_{\rm cont}(\vec{p}) &\equiv & (A-B_1 p_1^2-B_2 p_2^2) \sigma_x + C_1 p_1 \sigma_y +C_2 p_2 \sigma_z , \nonumber \\
\end{eqnarray}
where $A$, $B_1$, $B_2$, $C_1$, and $C_2$ are real constants. A similar argument as in the previous paragraph could be made to classify the Hamiltonian into two distinct topologies. It is thus expected that a discrete version of $\mathcal{H}^{(2D)}_{\rm cont}(p)$ similarly supports a 2D topological phase. Indeed, upon replacing $p_j \rightarrow \sin(k_j)$ and $p_j^2 \rightarrow 1-\cos(k_j)$ for $j=1,2$, we arrive at the QWZ model \cite{Qi06}, whose momentum space Hamiltonian can be written as (in dimensionless units)
\begin{eqnarray}
    \mathcal{H}_{\rm QWZ}(\mathbf{k})&=& (M-\cos(k_x)-\cos(k_y))\sigma_z +  \sin(k_x) \sigma_x \nonumber \\
    && + \sin(k_y) \sigma_y .
\end{eqnarray}
Since it breaks all (time-reversal, chiral, and particle-hole) symmetries, the QWZ model belongs to the A class in the AZ symmetry classification \cite{Altland97}, which is characterized by an integer topological invariant. For such a 2D topological system, the appropriate topological invariant is the so-called Chern number \cite{Jungwirth02}, defined as
\begin{equation}
    C_\ell = \frac{1}{2\pi} \int_{\rm BZ} \left( \mathrm{i} \langle \frac{\partial \Psi_\ell }{\partial k_x} | \frac{\partial \Psi_\ell }{\partial k_y} \rangle +h.c. \right) dk_x dk_y , \label{eq:chern}
\end{equation}
where $\ell=\pm$ is the band index, $|\frac{\partial \Psi_\ell}{\partial k_j}\rangle \equiv \frac{\partial }{\partial k_j } |\Psi_\ell\rangle$, $|\Psi_\ell\rangle$ is the $\ell$-valued energy eigenstate, and the integration is over the 2D Brillouin zone (BZ) $(-\pi,\pi]\times (-\pi,\pi]$. For the QWZ model, the Chern number of its two bands can be analytically computed as \cite{Qi06}
\begin{equation}
    C_\pm = \begin{cases}
        \pm 1 & \text{ for } 0<M<2 \\
        \mp 1 & \text{ for } -2<M<0 \\
        0 & \text{ for } |M|>2 \\
    \end{cases} .
\end{equation}
That is, the QWZ model is topologically trivial (nontrivial) for $|M|>2$ ($|M|<2$).

In the spirit of bulk--boundary correspondence, it can be explicitly checked that QWZ model supports a pair of topological edge modes in the topologically nontrivial regime. To this end, we first write the full (second-quantized) Hamiltonian of the \ch{QWZ} model as
\begin{equation}
    H_{\rm QWZ} = \sum_{k_x,k_y} \psi_{k_x,k_y}^\dagger \mathcal{H}_{\rm QWZ}(\mathbf{k}) \psi_{k_x,k_y} ,
\end{equation}
where $\psi_{k_x,k_y}=\left(c_{k_x,k_y,A},c_{k_x,k_y,B}\right)^T$ and $c_{k_x,k_y,S}$ is the particle annihilation operator at sublattice $S=A,B$ and quasimomenta $(k_x,k_y)$. It can be transformed to the appropriate real space operator according to \ch{
\begin{equation}
    c_{i,j,S}= \frac{1}{\sqrt{N_x N_y}}\sum_{k_x,k_y} c_{k_x,k_y,S} e^{-\mathrm{i} (k_x i+k_y j)},
\end{equation}
where the summation is over $k_x=-\pi+\frac{2\pi p}{N_x}$ and $k_y=-\pi+\frac{2\pi q}{N_y}$ for $p=1,2,\cdots, N_x$ and $q=1,2,\cdots,N_y$}, $N_x$ and $N_y$ are the lattice sizes in the $x$- and $y$-directions respectively. 

\begin{center}
    \begin{figure}
        \centering
        \includegraphics[scale=0.5]{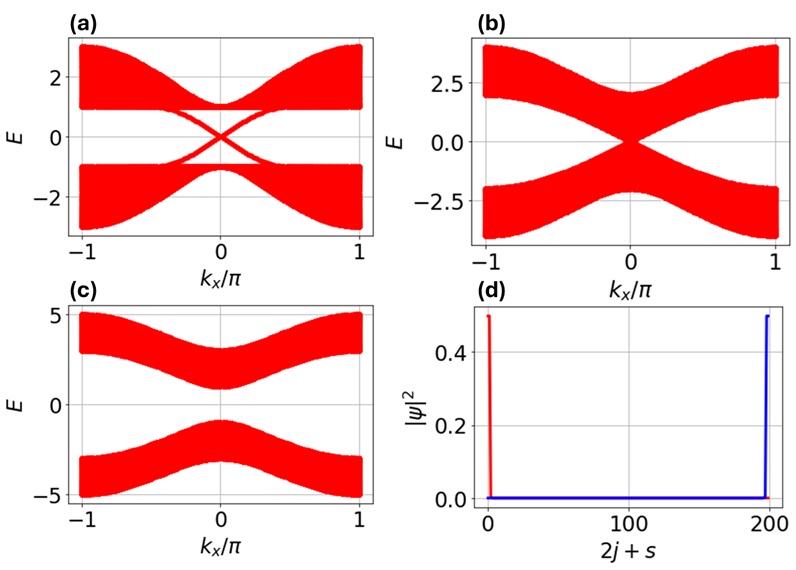}
        \caption{(a-c) The energy bands of $h_{QWZ}(k_x)$ at (a) $M=1$, (b) $M=2$, and (c) $M=3$. (d) The typical spatial profile (see main text for its definition) of the two edge modes of $h_{QWZ}(k_x)$ at $M=1$. The lattice size is taken as $N_y=100$ in all cases.}
        \label{fig:QWZ}
    \end{figure}
\end{center}

For the purpose of observing topological edge modes, it is also useful to define the ``mixed" space operators \ch{
\begin{equation}
    c_{k_x,j,S}= \frac{1}{\sqrt{N_y}}\sum_{k_y} c_{k_x,k_y,S} e^{-\mathrm{i} k_y j}.
\end{equation}
}Under OBC in the $y$-direction and PBC in the $x$-direction, the QWZ Hamiltonian can be written as
\begin{equation}
    H_{\rm QWZ} = \sum_{k_x} h_{QWZ}(k_x),
\end{equation}
where
\begin{widetext}
\begin{eqnarray}
    h_{QWZ}(k_x) &=& \sum_{j=1}^{N_y} \left[ (M-\cos(k_x)) c_{k_x,j,A}^\dagger c_{k_x,j,A} -(M-\cos(k_x)) c_{k_x,j,B}^\dagger c_{k_x,j,B} \right] +\sum_{j=1}^{N_y} \left[ \sin(k_x) c_{k_x,j,A}^\dagger c_{k_x,j,B} +h.c. \right] \nonumber \\
    && \sum_{j=1}^{N_y-1} \frac{1}{2} \left[c_{k_x,j+1,B}^\dagger c_{k_x,j,B}-c_{k_x,j+1,A}^\dagger c_{k_x,j,A}+ c_{k_x,j+1,B}^\dagger c_{k_x,j,A} -c_{k_x,j+1,A}^\dagger c_{k_x,j,B} +h.c.\right] .
\end{eqnarray}    
\end{widetext}
In the presence of only one particle in the 2D lattice, $h_{\rm QWZ}(k_x)$ can be easily diagonalized within the single particle subspace. Figure~\ref{fig:QWZ}(a)-(c) shows all of its eigenvalues as a function of $k_x$ for different values of $M$. As expected, a pair of chiral edge modes exist for $M=1$, as reflected by the in-gap band crossing in Fig.~\ref{fig:QWZ}(a). At exactly $M=2$, the two bands touch, signifying a topological phase transition. At larger values of $M$, such modes are absent, i.e., at $M=3$ in Fig.~\ref{fig:QWZ}(c). 
To verify the localization of the edge modes in the topologically nontrivial regime, we plot in Fig.~\ref{fig:QWZ}(d) the spatial profile of the corresponding edge modes in the topologically nontrivial regime. Here, we define the spatial profile $\psi$ as \ch{
\begin{equation}
    \psi(2j+s) = \langle j,s | E_{\rm edge} \rangle  , \label{eq:sprof}
\end{equation}
}where $|E_{\rm edge}\rangle$ is the single particle eigenstate of $h_{\rm QWZ}(k_x)$ (within the single particle subspace) associated with an edge mode and $|j,s\rangle$ represents a single particle basis state corresponding to a particle occupying site $j$ and sublattice $A$ ($B$) for $s=0$ ($s=1$). That the spatial profile of each edge mode exhibits a sharp peak either near $2j+s=0$ and \ch{$2j+s=2N_y$} in Fig.~\ref{fig:QWZ}(d) demonstrates its localization to one edge of the system, as expected. 

\subsubsection{Extension to 3D topological insulators}

The momentum space Hamiltonian $\mathcal{H}_{\rm cont}(p)$ can further be generalized to 3D according to
\begin{eqnarray}
    \mathcal{H}^{(3D)}_{\rm cont}(\vec{p}) & \equiv & (A-B_1 p_1^2-B_2 p_2^2-B_3 p_3^2) \Gamma_1 + C_1 p_1 \Gamma_2 \nonumber \\
    && +C_2 p_2 \Gamma_3 + C_3 p_3 \Gamma_4,
\end{eqnarray}
where $\Gamma_j$ are the mutually anticommuting $\Gamma$ matrices \cite{Dirac28} that generalize the Pauli matrices. The corresponding lattice Hamiltonian is easily obtained as
\begin{eqnarray}
    \mathcal{H}_{\rm 3DTI} (\mathbf{k}) &=& (M-\cos(k_x)-\cos(k_y)-\cos(k_z))\Gamma_1 \nonumber \\ 
    && +  \sin(k_x) \Gamma_2 + \sin(k_y) \Gamma_3 + \sin(k_z) \Gamma_4 \nonumber , \\ \label{eq:qwz3d} 
\end{eqnarray}
which describes a 3D topological insulator as elaborated below.

Without loss of generality, we may write the four $\Gamma$ matrices in terms of the Kronecker product between two Pauli matrices as $\Gamma_j=\sigma_y \otimes \sigma_j$ for $j=1,2,3$ and $\Gamma_4=\sigma_z \otimes \mathcal{I}_2$. We can then identify the time-reversal, particle-hole, and chiral symmetry operators as $\mathcal{T}=\mathcal{K} (\mathrm{i} \sigma_y) \otimes \sigma_z$, $\mathcal{P}=\mathcal{K} \sigma_z \otimes \sigma_z $, and $\mathcal{C}=\sigma_x\otimes \mathcal{I}_2$. Moreover, as $\mathcal{T}^2=-\mathcal{P}^2=-1$, the system belongs to the DIII class in the AZ symmetry classification \cite{Altland97}.

While a system in the DIII class generally supports an integer topological invariant \cite{Altland97}, in the literature, such a 3D topological insulator is usually characterized by a $\mathbb{Z}_2$ invariant instead. The advantage of such a description is that it allows for the presence of perturbations that break all but time-reversal symmetries. Indeed, a 3D topological insulator protected by a time-reversal symmetry alone with $\mathcal{T}^2=-1$ belongs to the AII class that is characterized by a $\mathbb{Z}_2$ invariant \cite{Altland97}. 

To construct the appropriate $\mathbb{Z}_2$ invariant, it is first worth noting the presence of eight time-reversal invariant momenta (TRIM) $(k_x,k_y,k_z)=(\kappa_x,\kappa_y,\kappa_z)$ with $\kappa_x,\kappa_y,\kappa_z\in\left\lbrace 0,\pi \right\rbrace$, at which $\mathcal{H}_{\rm 3DTI} (\mathbf{k}=\mathbf{\kappa})$ and $\mathcal{T}$ commute. At these points, the eigenvalues of $\mathcal{H}_{\rm 3DTI} (\mathbf{k}=\mathbf{\kappa})$ are simply the doubly-degenerate energies
\begin{equation}
    E_{\rm 3DTI,\pm} (\mathbf{k}=\mathbf{\kappa}) = \pm (M-\cos(\kappa_x)-\cos(\kappa_y)-\cos(\kappa_z)) ,  
\end{equation}
which correspond to $\sigma_y \otimes \sigma_x =\pm 1$ respectively. In particular, depending on the sign of $(M-\cos(\kappa_x)-\cos(\kappa_y)-\cos(\kappa_z))$, there may be a sign difference between $E_{\rm 3DTI,\pm} (\mathbf{k}=\mathbf{\kappa})$ and $\sigma_y \otimes \sigma_x$. For each TRIM, we can then define the corresponding eigenstates' parity as 
\begin{equation}
    P(\vec{\kappa})=\mathrm{sgn}\left(M-\cos(\kappa_x)-\cos(\kappa_y)-\cos(\kappa_z)\right) .   
\end{equation}
The system's $\mathbb{Z}_2$ invariant could then be obtained as the product of the eigenstates' parities at all TRIM, i.e.,
\begin{equation}
    (-1)^\nu \equiv \prod_{\vec{\kappa}} P (\vec{\kappa}) = \mathrm{sgn}\left[(M^2-9)(M^2-1)\right] .
\end{equation}
Intuitively, the invariant $\nu$ indirectly predicts the presence/absence of nontrivial twists in the eigenstates' structure over the 3D BZ, which in turn dictates the topology. The case $\nu=0$ is topologically equivalent to the case in which there is no twist, i.e., $P(\vec{\kappa})$ is equal at all TRIM, thereby corresponding to the trivial case. Meanwhile, $\nu=1$ corresponds to the topologically nontrivial regime. That is, our 3D topological insulator is topologically nontrivial for $1<|M|<3$ and is topologically trivial otherwise.

\begin{center}
    \begin{figure*}
        \centering
        \includegraphics[scale=0.5]{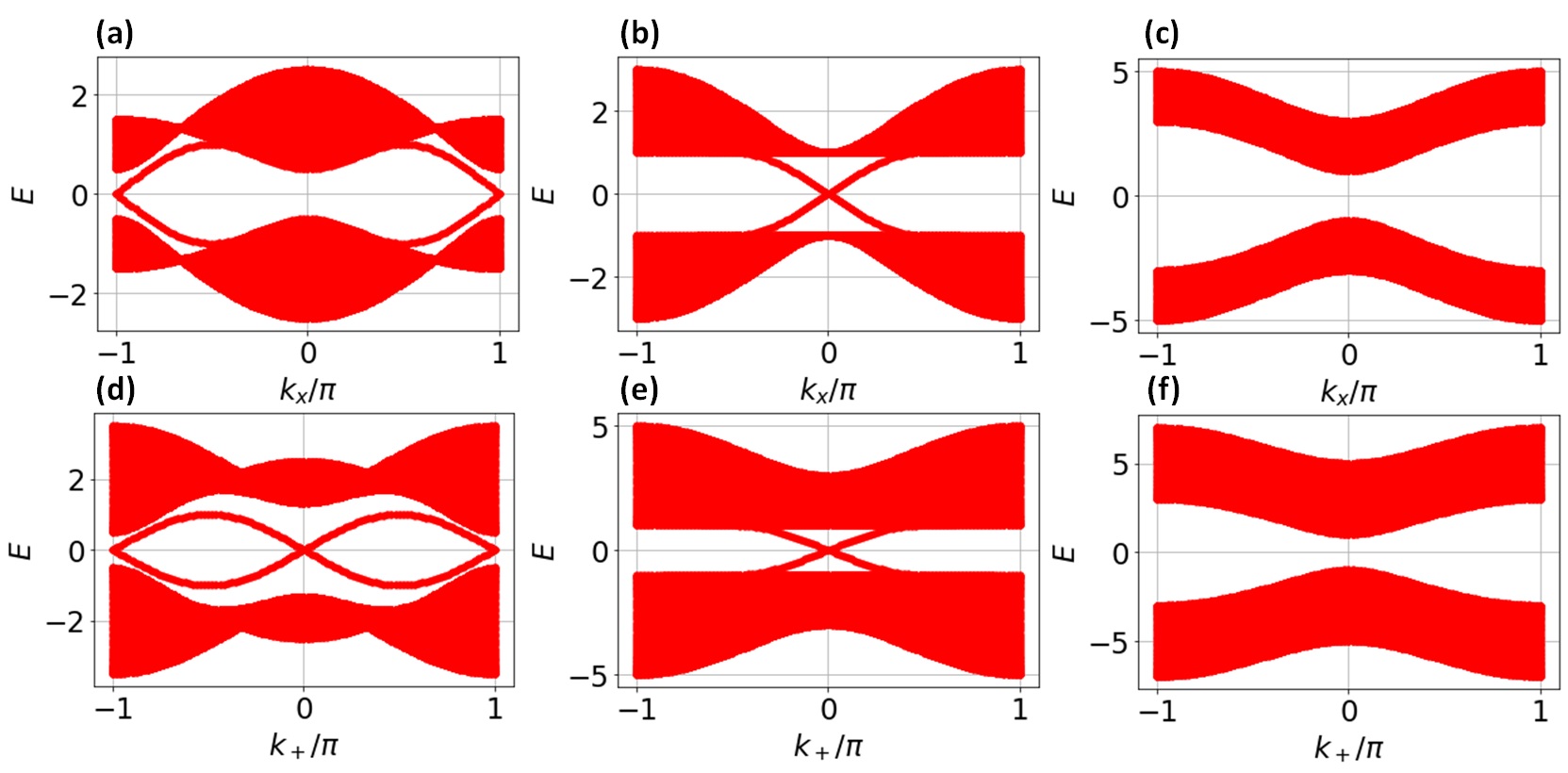}
        \caption{The energy band structure of the 3D topological insulating model of Eq.~(\ref{eq:qwz3d}) within the single particle subspace at a fixed $k_z=0$. In panels (a)-(c), OBC is taken along the $y$-direction, whereas in panels (d)-(f), OBC is taken along the diagonal $(x+y)$-direction. The parameter $M$ is taken as (a,d) $M=0.5$, (b,e) $M=2$, (c,f) $M=4$, whilst the lattice size is taken as (a,b,c) $N_y=50$ and (d,e,f) $N_-=50$. }
        \label{fig:3DTIspec}
    \end{figure*}
\end{center}

Under OBC in the $y$-direction and PBC in the $x$- and $z$-direction, diagonalizing the corresponding second-quantized Hamiltonian within the single particle subspace in the spirit of the previous section yields the energy bands shown in Fig.~\ref{fig:3DTIspec}(a)-(c). As expected, \ch{in-gap boundary modes (corresponding to helical surface modes)}  are observed in the topologically nontrivial regime $1<|M|<3$, i.e., Fig.~\ref{fig:3DTIspec}(b), which are otherwise absent in the trivial regime $|M|>3$, i.e., Fig.~\ref{fig:3DTIspec}(c). 

Interestingly, \ch{similar-looking boundary modes} are also observed in the regime $|M|<1$ (Fig.~\ref{fig:3DTIspec}(a)), which was predicted to be topologically trivial by the $\mathbb{Z}_2$ invariant $\nu$. In fact, these boundary modes originate from the topology of a 2D system. Indeed, by fixing $k_z=0$, the second-quantized Hamiltonian reduces to the corresponding QWZ Hamiltonian discussed in the previous section with $M_{\rm eff} = M-1$. The topologically nontrivial regime of QWZ ($|M_{\rm eff}|<2$) then precisely translates to $|M|<3$, i.e., the $|M|<1$ region is also included. 

In the literature, the true topologically nontrivial regime ($1<|M|<3$) that yields $\nu=1$ corresponds to a ``strong topological insulator" (STI), whilst the topologically trivial regime that exhibits remnants of some lower dimensional topology ($|M|<1$) gives rise to a ``weak topological insulator" (WTI) \cite{Fu07}. Being a genuine 3D topological phase, boundary modes that emerge in an STI are typically insensitive to how the boundaries are introduced. By contrast, a WTI may or may not support boundary modes; even if it does, the profile strongly depends on the detail of the boundaries.  

To demonstrate the effect of the boundaries on the system's boundary modes, we consider a different geometry in which PBC is applied in the $z$- and the ``diagonal" $(x-y)$-direction, while OBC is applied in the other diagonal $(x+y)$-direction. By denoting $k_+$ as the quasimomentum in the $(x-y)$-direction, the band structure at different $M$ values is presented in Fig.~\ref{fig:3DTIspec}(d)-(f). In the WTI regime (Fig.~\ref{fig:3DTIspec}(d)), counter-propagating boundary modes are now observed in the $k_+$ direction. This is to be compared with the unidirectional boundary modes that emerge along the $k_x$ direction (Fig.~\ref{fig:3DTIspec}(d)). Such counter-propagating boundary modes are usually only deemed \ch{weakly protected}, as appropriate perturbations easily hybridize pairs of boundary modes moving in the opposite directions, which end up removing them. By contrast, in the STI regime (Fig.~\ref{fig:3DTIspec}(e)), \ch{helical surface modes} and qualitatively the same band structure as in Fig.~\ref{fig:3DTIspec}(b) are obtained, as expected. Finally, in the topologically trivial regime (Fig.~\ref{fig:3DTIspec}(f)), no boundary modes are observed, as it should be. 

\subsubsection{Extension to 3D Weyl semimetals}

Without resorting to $\Gamma$ matrices, a variation of $\mathcal{H}_{\rm 3DTI}(\vec{k})$ considered in the previous section yields 
\begin{eqnarray}
    \mathcal{H}_{\rm WSM}(\mathbf{k})&=& (M-\cos(k_x)-\cos(k_y)-\cos(k_z))\sigma_z  \nonumber \\
    && +  \sin(k_x) \sigma_x + \sin(k_y) \sigma_y . \label{eq:WSM}
\end{eqnarray}
At $|M|<3$, there are exactly two points $\vec{\kappa}_{\pm}$ at which the two bands of $\mathcal{H}_{\rm WSM}(\mathbf{k})$ touch at zero energy. Without loss of generality, we shall further constrain $M$ to satisfy $1<M<3$, so that the band touching points occur at $\vec{\kappa}_{\pm}=(0,0,\pm \arccos(M-2))$. In the vicinity of these two points, $\mathcal{H}_{\rm WSM}(\mathbf{k})$ reduces to the (continuum) massless \ch{Weyl} Hamiltonian. Consequently, the energy eigenvalues exhibit linear dispersion in all three quasimomenta near any of these two points, i.e.,
\begin{equation}
    E_\pm =\pm \sqrt{\delta k_x^2 + \delta k_y^2 +\delta k_z^2} .
\end{equation}
That is, the two band touching points correspond to 3D Dirac cones (often termed Weyl points). The Hamiltonian $\mathcal{H}_{\rm WSM}(\mathbf{k})$, which was first introduced in Ref.~\cite{Yang11}, thus represents a minimal model for the Weyl semimetal \cite{Wan11}. 

\ch{Before proceeding, it is worth emphasizing that the above description corresponds to an isotropic Weyl semimetal in which the velocity components are equal in all three directions. The ``anisotropic" version of the Weyl semimental can be obtained through $\sigma_x \rightarrow v_x \sigma_x$, $\sigma_y \rightarrow v_y \sigma_y$, and $\sigma_z \rightarrow v_z \sigma_z$ in Eq.~(\ref{eq:WSM}) with $v_x\neq v_y \neq v_z$. Nevertheless, both isotropic and anisotropic Weyl semimetals exhibit linear dispersion and share the same topological characterizations. Therefore, we shall focus on the former in the following discussion for simplicity.}

\begin{center}
    \begin{figure}
        \centering
        \includegraphics[scale=0.5]{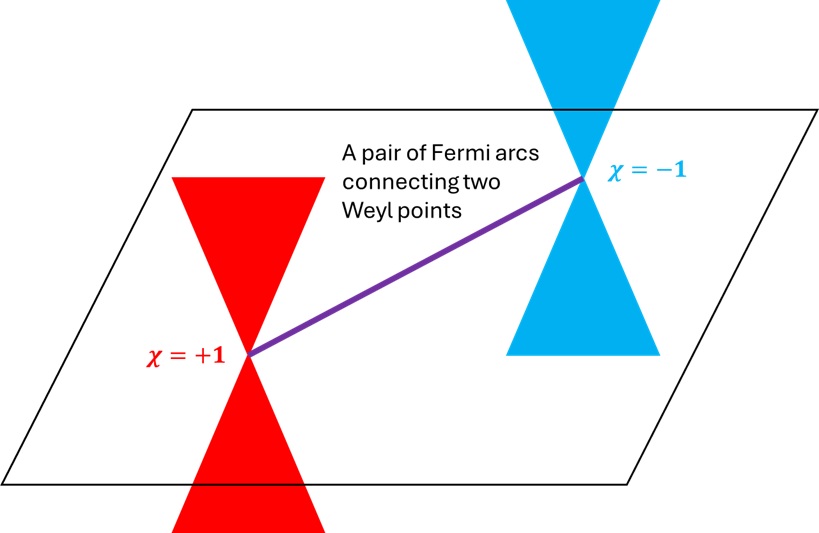}
        \caption{Schematic depiction of Weyl points with opposite chirality and the Fermi arcs connecting them. \ch{Note that only a portion of The Fermi arcs that intersects the appropriate 2D Brillouin zone is drawn.}}
        \label{fig:wsmdep}
    \end{figure}
\end{center}

Despite the lack of band gap, Weyl semimetal also represents a topological phase as it supports a pair of \ch{boundary} modes that are of topological origin. Specifically, such \ch{boundary} modes, which are usually termed the Fermi arcs, \ch{represent open contours connecting the projections of a pair of Weyl points onto the surface of quasimomenta (see Fig.~\ref{fig:wsmdep}) that remain good quantum numbers (2D Brillouin zone). Despite the name, the Fermi arcs do not strictly form dispersionless arcs, except along a specific quasimomentum direction and at a fixed energy slice. The Fermi arcs may otherwise disperse in energy along a generic direction. The topological nature of the Fermi arcs stems from the fact that they connect a pair of Weyl points with opposite ``chirality" ($\chi$).} Here, the chirality of a Weyl point can be computed by creating a small sphere in the momentum space that encloses the Weyl point and evaluating the Chern number of one band over the surface of the sphere.   

To illustrate the above procedure for calculating the chirality of a Weyl point, suppose that its low energy Hamiltonian takes the form
\begin{equation}
    \mathcal{H}_{\rm low}(\mathbf{k}) = k_x \sigma_x +k_y \sigma_y + k_z \sigma_z .
\end{equation}
A sphere surrounding the Weyl point can be made by parameterizing $k_x=r \cos(\phi)\sin(\theta)$, $k_y=r \sin(\theta) \sin(\phi)$, and $k_z= r \cos(\theta)$, where $r$ is the radius of the sphere, $\theta\in[0,\pi)$, and $\phi\in[0,2\pi)$. One eigenstate of $\mathcal{H}_{\rm low}$ can be written as
\begin{equation}
    |\Psi_+\rangle = \left(\cos(\theta/2),\sin(\theta/2)e^{\mathrm{i}\phi}\right)^T . 
\end{equation}
Using Eq.~(\ref{eq:chern}), the chirality of the Weyl point can be explicitly calculated through direct integration, i.e.,
\begin{equation}
    \chi = \frac{1}{2\pi} \int_{0}^{2\pi} \int_0^{\pi} \frac{\sin\theta}{2} d\theta d\phi = +1 .
\end{equation}

As the Hamiltonian describing a Weyl semimetal involves all Pauli matrices $\sigma_x$, $\sigma_y$, and $\sigma_z$, a generic perturbation that preserves the number of degrees of freedom cannot destroy its Weyl points, except through pairwise annihilation of Weyl points with opposite chiralities when they coincide in the momentum space. In addition to this robustness, significant interest towards Weyl semimetals also stems from their unusual transport properties such as negative magnetoresistance \cite{Niel83,Aji12,Kim13}, anomalous Hall effect \cite{Burkov11,Xu11,Zyuzin12}, and the chiral magnetic
effect \cite{Zyuzin12b,Taguchi16}.

\subsection{SSH models as building blocks of higher-order topological phases}
\label{HOSSH}

As elucidated in the previous section, extending the momentum space Hamiltonian of the SSH model to a higher dimension yields a variety of higher-dimensional topological phases ranging from 2D topological insulators to Weyl semimetals. Now, we turn our attention to the real space Hamiltonian of the SSH model, which follows the structure of Fig.~\ref{fig:sshdep}. One of its most natural extensions to a 2D setting is to simply consider multiple copies of the SSH Hamiltonians and couple them in a particular manner. 

\begin{center}
    \begin{figure}
        \centering
        \includegraphics[scale=0.6]{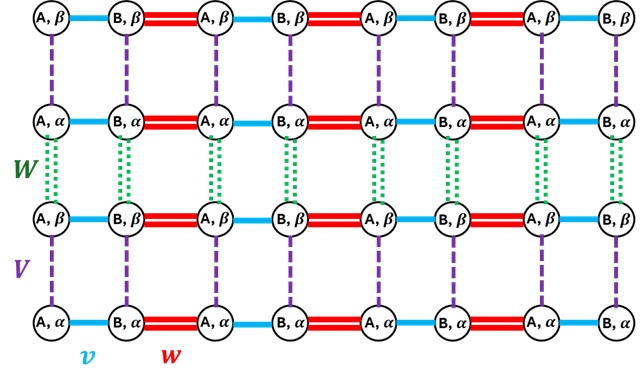}
        \caption{Schematic description of a stack of SSH chains under an alternating coupling scheme.}
        \label{fig:stidep}
    \end{figure}
\end{center}

Given that the seemingly simple alternating nearest-neighbor coupling structure suffices to yield nontrivial topology, it is instructive to apply the same coupling structure on a stack of SSH chains, as schematically shown in Fig.~\ref{fig:stidep}. The resulting 2D Hamiltonian then reads 

\begin{widetext}
\begin{eqnarray}
    H_{\rm SSH-SSH} &=& \sum_{\ell=\alpha,\beta} \left[\sum_{(i,j)=(1,1)}^{(N_x,N_y)} v_\ell c_{i,j,B,\ell}^\dagger c_{i,j,A,\ell} + \sum_{(i,j)=(1,1)}^{(N_x-1,N_y)} w_\ell c_{i+1,j,A,\ell}^\dagger c_{i,j,B,\ell} +h.c. \right] \nonumber \\
    && +\sum_{S=A,B} \left[\sum_{(i,j)=(1,1)}^{(N_x,N_y)} V_S c_{i,j,S,\beta}^\dagger c_{i,j,S,\alpha} + \sum_{(i,j)=(1,1)}^{(N_x,N_y-1)} W_S c_{i,j+1,S,\alpha}^\dagger c_{i,j,S,\beta} +h.c.\right] , \label{eq:sotissh}
\end{eqnarray}    
\end{widetext}
where $v_\ell$ and $w_\ell$ for $\ell=\alpha,\beta$ are the hopping parameters within each SSH chain, $V_S$ and $W_S$ for $S=A,B$ are the hopping parameters between neighboring 1D chains, $c_{i,j,S,\ell}$ ($c_{i,j,S,\ell}^\dagger$) is the corresponding particle annihilation (creation) operator at site $(i,j)$ and sublattice $(S,\ell)$, $N_x$ and $N_y$ are respectively the length of each SSH chain and the length of the stack. Note that to keep the model general, $v_\ell$ and $w_\ell$ ($V_S$ and $W_S$) are allowed to take distinct values at $\ell=\alpha$ and $\ell=\beta$ ($S=A$ and $S=B$).

The model $H_{SSH-SSH}$ is known to support a second-order topological phase \cite{Benalcazar17,Benalcazar17b}, which is marked by the presence of zero energy modes at its corners, i.e., boundaries of its boundaries. Indeed, the emergence of zero energy corner modes in this specific model could be intuitively understood by considering the special parameter values $v_\ell=V_S=0$, at which the four operators $c_{1,1,A,\alpha}$, $c_{N_x,1,B,\alpha}$, $c_{1,N_y,A,\beta}$, and $c_{N_x,N_y,B,\beta}$ commute with $H_{SSH-SSH}$. Away from these special parameter values, zero energy corner modes also exist over a window of $v_\ell$, $w_\ell$, $V_S$, and $W_S$ values, which are further backed by some appropriate topological invariant \cite{Benalcazar17,Benalcazar17b}.

The interpretation of $H_{SSH-SSH}$ as a stack of SSH chains was highlighted in Ref.~\cite{Bomantara19}. There, $v_\alpha=v_\beta=v$, $w_\alpha=w_\beta$, $V_A=V_B=V$, and $W_A=W_B=W$ are taken for further simplification. The corresponding momentum space Hamiltonian can then be written as\ch{
\begin{eqnarray}
    \mathcal{H}_{SSH-SSH}(\mathbf{k}) &=& \left(v+w\cos(k_x)\right) \sigma_x + w \sin(k_x) \sigma_y \nonumber \\
    &+& \left(V+W\cos(k_y)\right) \tau_x + W \sin(k_y) \tau_y , \nonumber \\ \label{eq:sotirwb}
\end{eqnarray}
}where $\sigma$ and $\tau$ are Pauli matrices acting on the intra- and inter-chain sublattices respectively. Unlike first-order topological phases, second-order topological phases are typically protected by a combination of internal (chiral, particle-hole, and/or time-reversal) symmetries and spatial symmetries, e.g., inversion and mirror symmetries \cite{Benalcazar17,Benalcazar17b}. Indeed, for the case of $H_{SSH-SSH}$, it not only possesses chiral, particle-hole, and time-reversal symmetries according to $\mathcal{C}=\sigma_z\tau_z$, $\mathcal{P}=\mathcal{K}\sigma_z\tau_z$, and $\mathcal{T}=\mathcal{K}$ respectively, but the system also exhibits inversion symmetry $\mathcal{I}= \sigma_x\tau_x$, as well as two reflection symmetries $\mathcal{M}_x=\sigma_x$ and $\mathcal{M}_y=\tau_x$. These spatial symmetries satisfy
\begin{eqnarray}
    \mathcal{I}^{-1} \mathcal{H}_{SSH-SSH}(\mathbf{k}) \mathcal{I} &=& \mathcal{H}_{SSH-SSH}(-\mathbf{k}) , \nonumber \\
    \mathcal{M}_x^{-1} \mathcal{H}_{SSH-SSH}(\mathbf{k}) \mathcal{M}_x &=& \mathcal{H}_{SSH-SSH}(-k_x,k_y) , \nonumber \\
    \mathcal{M}_y^{-1} \mathcal{H}_{SSH-SSH}(\mathbf{k}) \mathcal{M}_y &=& \mathcal{H}_{SSH-SSH}(k_x,-k_y) . \nonumber \\
\end{eqnarray}

As Eq.~(\ref{eq:sotirwb}) is a Kronecker sum between two SSH Hamiltonians, its topological invariant can be obtained by taking the product of the winding numbers of each SSH Hamiltonian. Specifically, 
\begin{equation}
    \mathcal{W}_{xy} = \mathcal{W}_x \times \mathcal{W}_y , 
\end{equation} 
where 
\begin{eqnarray}
    \mathcal{W}_x= \frac{1}{2\pi \mathrm{i}} \oint_{\mathcal{C}_x} \frac{1}{v+z} dz &,& \mathcal{W}_y = \frac{1}{2\pi \mathrm{i}} \oint_{\mathcal{C}_y} \frac{1}{V+z} dz , \nonumber \\
\end{eqnarray}
$\mathcal{C}_x$ and $\mathcal{C}_y$ are respectively circular contours of radii $w$ and $W$ centered around the origin in the complex plane. By analytically computing these integrals, one obtains
\begin{equation}
    \mathcal{W}_{xy} = \begin{cases}
        1 & \text{ for }w>v\text{ and }W>V \\
        0 & \text{otherwise}
    \end{cases} .
\end{equation}

\begin{center}
    \begin{figure}
        \centering
        \includegraphics[scale=0.35]{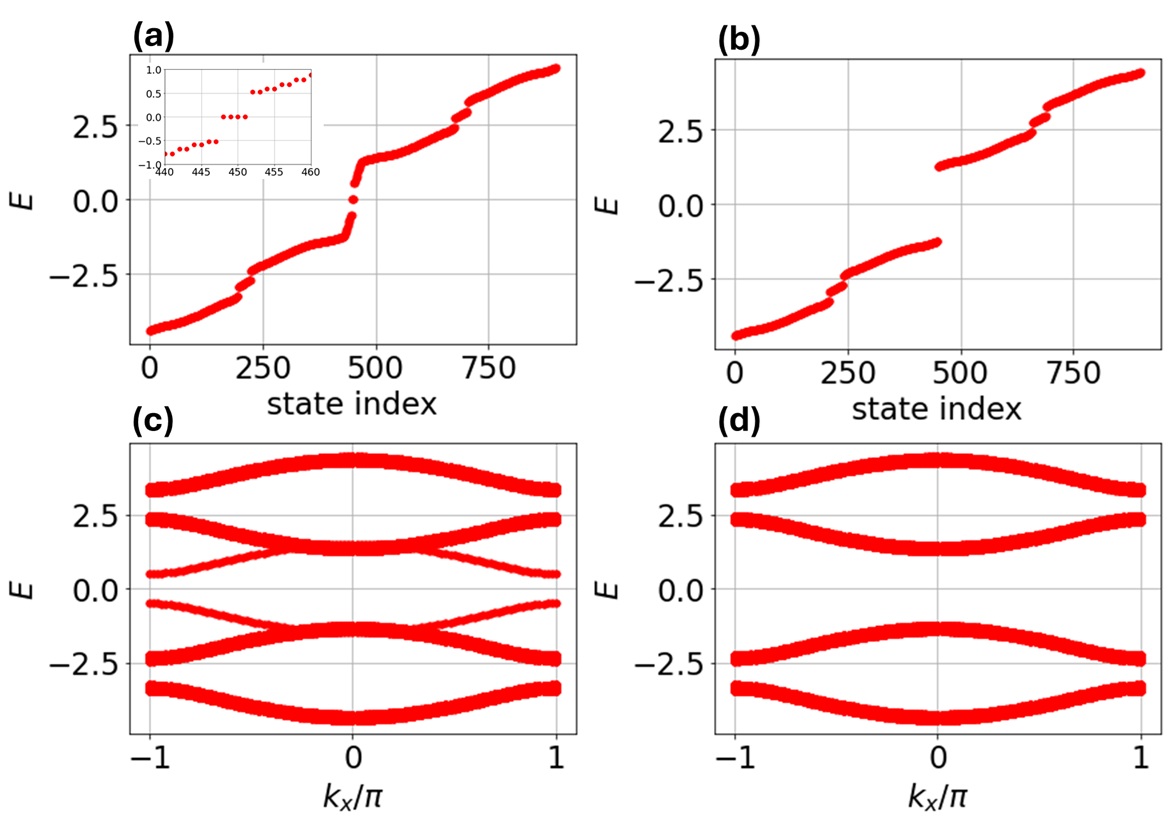}
        \caption{The single particle energy spectrum of Eq.~(\ref{eq:sotissh}) under (a,b) OBC in both $x$- and $y$-directions, (c,d) OBC in the $y$-direction and PBC in the $x$-direction. The inset of panel a shows the zoomed-in view of the energy spectrum in the vicinity of the zero energy corner modes. The system parameters are taken as $v=0.5$, $w=1$, (a,c) $V=0.1$, $W=2.85$, (b,d) $V=2.85$, $W=0.1$. The system size is taken as $(N_x,N_y)=(15,15)$ in panels (a,b) and $N_y=50$ in panels (c,d).}
        \label{fig:soti}
    \end{figure}
\end{center}

The presence of zero energy corner modes in the topologically nontrivial regime of $\mathcal{W}_{xy}=1$ can be explicitly verified by diagonalizing Eq.~(\ref{eq:sotissh}) within the single particle subspace, which yields the energy eigenvalues shown in Fig.~\ref{fig:soti}. In particular, four zero energy eigenvalues are obtained when $\mathcal{W}_{xy}=1$ and OBC are applied to both $x$- and $y$-directions, i.e., see Fig.~\ref{fig:soti}(a) and its inset for a clearer view of the zero energy eigenvalues. These zero energy eigenvalues are otherwise absent if the system is in the topologically trivial regime with $\mathcal{W}_{xy}=0$ (Fig.~\ref{fig:soti}(b)). Moreover, regardless of whether the system is topologically trivial or nontrivial, zero energy eigenvalues are also absent if no corner is present, even if there is an edge. This is evidenced from Fig.~\ref{fig:soti}(c,d), which show the energy band structure in the topologically nontrivial and trivial regimes respectively under OBC in the $y$-direction and PBC in the other. In particular, the additional gapped bands observed in Fig.~\ref{fig:soti}(c) correspond to edge states. Therefore, the four zero energy eigenvalues observed in Fig.~\ref{fig:soti}(a) indeed correspond to topological corner modes.

\section{Extending the SSH model by modifying its sublattice structure}

By definition, the SSH model consists of two sublattices. By simply enlarging its sublattice size while otherwise maintaining the remaining structure intact, the resulting model has been found to exhibit significantly richer properties. In existing studies, various such extensions have been made from different starting perspectives of the SSH chain, i.e., its real space or momentum space model. 

\subsection{Extending the SSH model from its real space Hamiltonian}
\label{SSH3}

\subsubsection{Extension by changing the periodicity of the hopping amplitudes}

\begin{center}
    \begin{figure}
        \centering
        \includegraphics[scale=0.7]{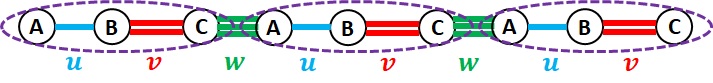}
        \caption{Schematic description of the extended SSH Hamiltonian with $N=3$. Each unit cell (dashed ellipse) consists of three sublattices labeled as A, B, and C.}
        \label{fig:ssh3dep}
    \end{figure}
\end{center}

As was schematically shown in Fig.~\ref{fig:sshdep}, the real space Hamiltonian of the SSH model consists of two alternating hopping amplitudes $v$ and $w$. One of the most straightforward extensions of this structure is to consider a Hamiltonian with more than two hopping amplitudes. For example, in Ref.~\cite{Alv19}, an extended SSH model with three alternating hopping amplitudes $u$, $v$, and $w$ (as depicted in Fig.~\ref{fig:ssh3dep}) are proposed and investigated. The Hamiltonian describing such a model reads
\begin{equation}
    H= \sum_{j=1}^{N} \left(u c_{j,B}^\dagger c_{j,A} + v c_{j,C}^\dagger c_{j,B}  \right)+ \sum_{j=1}^{N-1} w c_{j+1,A}^\dagger c_{j,C} +h.c., \label{ssh3mod}
\end{equation}
the corresponding momentum space Hamiltonian of which is (if PBC are assumed)
\begin{equation}
    \mathcal{H}(k)= \left( \begin{array}{ccc}
       0 & u & w e^{-\mathrm{i} k}  \\
       u & 0 & v \\
       w e^{\mathrm{i} k} & v & 0 \\
    \end{array} \right) . \label{ssh3modmom}
\end{equation}
As $\mathcal{H}(k)$ is $3\times 3$, diagonalizing Eq.~(\ref{ssh3modmom}) as a function of $k$ yields three energy bands instead of two as in the regular SSH chain. For this reason, such a three-band extended SSH model is often also termed the SSH3 model \cite{Du24}.  

\begin{center}
    \begin{figure}
        \centering
        \includegraphics[scale=0.37]{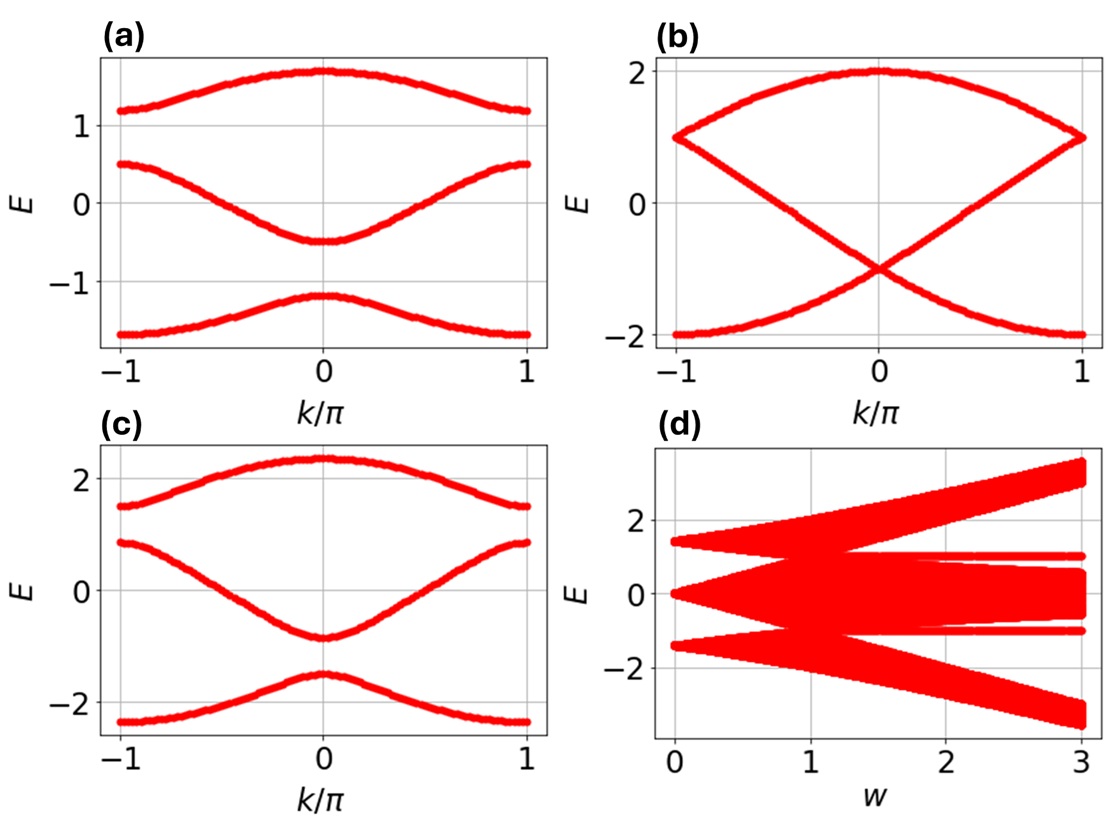}
        \caption{The single particle energy spectrum of the SSH3 model under (a,b,c) PBC and (d) OBC at varying values of $w$. The system parameters are taken as $u=v=1$ (all panels) and (a) $w=0.5$, (b) $w=1$, (c) $w=1.5$. In (d), a system size of $N=100$ is chosen.}
        \label{fig:ssh3spec}
    \end{figure}
\end{center}

\begin{center}
    \begin{figure}
        \centering
        \includegraphics[scale=0.42]{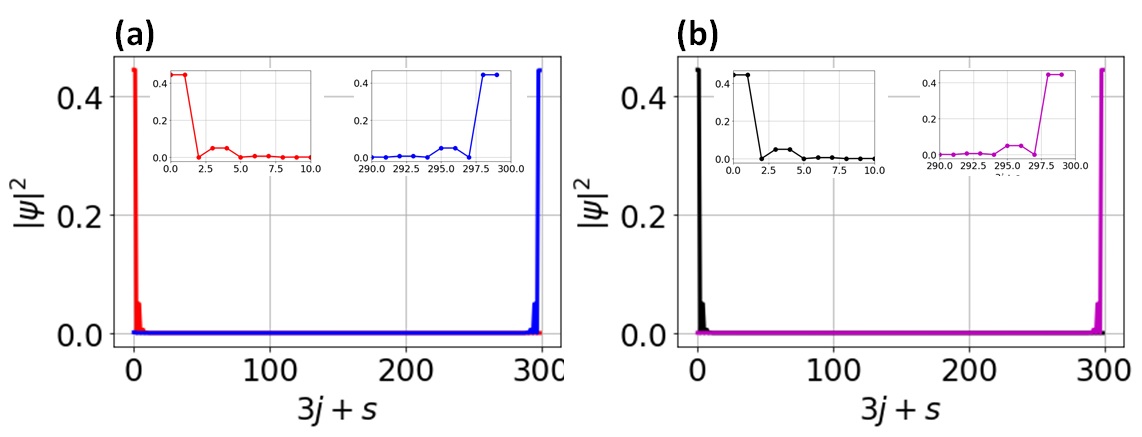}
        \caption{The spatial profile of the four edge modes in the SSH3 model at $u=v=1$ and $w=3$. Panel (a) corresponds to edge modes with positive energies, whereas panel (b) corresponds to edge modes with negative energies. In both panels, the insets depict the zoomed-in view of each edge mode near its main peaks.}
        \label{fig:ssh3prof}
    \end{figure}
\end{center}

Figure~\ref{fig:ssh3spec}(a-c) depicts the band structure of the SSH3 model under PBC for 3 different values of intercell hopping amplitudes $w$, while Fig.~\ref{fig:ssh3spec}(d) shows the corresponding single particle energy spectrum under OBC as a function of $w$. Under PBC, it is observed that, similarly to the regular SSH chain, the bands become gapless when the three hopping amplitudes are equal, i.e., $u=v=w$, signifying a topological phase transition. Away from these special parameter values, the three bands are generally gapped and look qualitatively the same, i.e., Fig.~\ref{fig:ssh3spec}(a,c). However, in the presence of edges (when OBC are applied), the regime $w>u,v$ is distinguished from the regime $w<u,v$ by the presence of additional edge localized but nonzero energy solutions, i.e., Fig.~\ref{fig:ssh3spec}(d). In Fig.~\ref{fig:ssh3prof}, the spatial profile of all edge modes in the SSH3 model are presented, where a point $3j+s$ corresponds to site $j$ and sublattice A, B, or C for $s=0,1$, and $2$ respectively. 

Unlike the regular SSH model, the SSH3 lacks chiral symmetry. However, in the special case of $u=v$, the system respects the inversion symmetry, i.e., $\mathcal{I}^{-1}\mathcal{H}(k)\mathcal{I}=\mathcal{H}(k)$ under \cite{Alv19}
\begin{equation}
    \mathcal{I} = \left( \begin{array}{ccc}
        0 & 0 & 1 \\
        0 & 1 & 0 \\
        1 & 0 & 0
    \end{array} \right) .
\end{equation}
The Zak phase (as previously defined in Eq.~(\ref{Eq:zak})) then continues to take a quantized value of either $\pi$ or $0$ and can serve as the system's topological invariant. In particular, it can be analytically verified that the Zak phase corresponding to the lower band of the SSH3 model is $\pi$ ($0$) for $w>u=v$ ($w<u=v$) \cite{Alv19}.

At more general parameter values of $u$, $v$, and $w$, the SSH3 model no longer respects the inversion symmetry. However, as identified in Ref.~\cite{Anas22}, it actually respects a type of extended chiral symmetry, termed \emph{point chirality}, that satisfies
\begin{equation}
\Gamma_p^{-1} \mathcal{H}(k)\Gamma_p = \Gamma_p^{-1} \mathcal{H}^*(-k)\Gamma_p = -\mathcal{H}^*(\pi-k) ,  
\end{equation}
where $\Gamma_p=\mathrm{diag}(1,-1,1)$. Intuitively, if the energy bands of the SSH3 are arranged so that $E_0(k)<E_1(k)<E_2(k)$, the point chiral symmetry maps the set of energy eigenvalues $(E_0(k),E_1(k),E_2(k))$ to $(-E_2(\pi-k),-E_1(\pi-k),-E_0(\pi-k))$. That is, within the right half of the Brillouin zone $[0,\pi)$, the energy bands are symmetric about $k=\pi/2$ and $E=0$ (in the left half of the Brillouin zone, $[-\pi,0)$, the energy bands are similarly symmetric about $k=-\pi/2$ and $E=0$). Following Ref.~\cite{Anas22}, the regime $[0,\pi)$ shall be referred to as the ``reduced Brillouin zone".

The lack of both chiral and inversion symmetry makes the regular Zak phase unsuitable to serve as a topological invariant, since it is now not constrained to take a quantized value. In view of this, Ref.~\cite{Anas22} proposes a modification to the Zak phase, termed ``normalized sublattice Zak phase", that incorporates the point chiral symmetry above as a possible topological invariant for the SSH3 model. It is defined as
\begin{equation}
    Z^\lambda_{A,C} \equiv \frac{\mathrm{i}}{2} \oint \langle \tilde{u}_\lambda(k)|\partial_k \tilde{u}_\lambda(k) \rangle dk , \label{eq:nszak}
\end{equation}
where
\begin{equation}
    | \tilde{u}_\lambda(k) \rangle \equiv \frac{P_A |u_\lambda (k)\rangle}{\sqrt{\langle u_\lambda (k) | P_A | u_\lambda (k)\rangle}} ,
\end{equation}
$|u_\lambda (k)\rangle$ is the eigenstate corresponding to the band index $\lambda=0,1,2$, and $P_A$ is the projector onto the A sublattice. The second subscript in Eq.~(\ref{eq:nszak}) denotes the gauge choice of the eigenstate, i.e., ``C" means that the third component of the eigenstate (corresponding to sublattice C) is taken as a real number.

\begin{center}
    \begin{figure}
        \centering
        \includegraphics[scale=0.42]{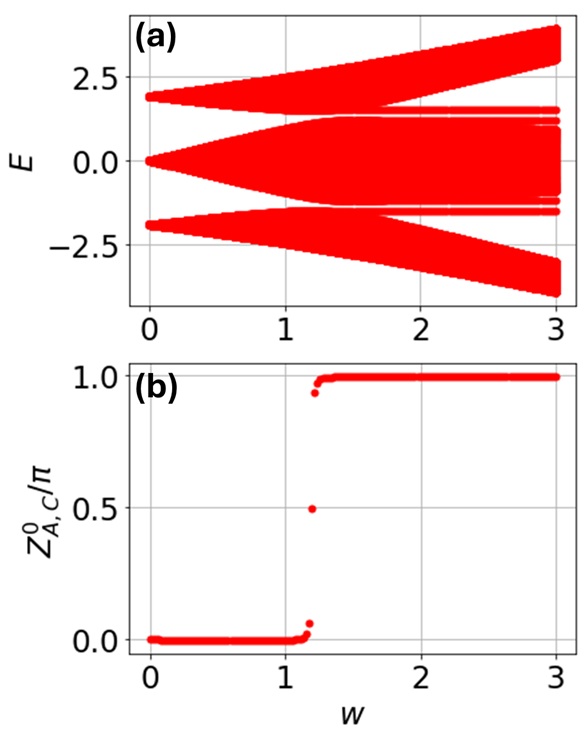}
        \caption{(a) The single particle energy spectrum of the SSH3 model under OBC at $N=100$. (b) The numerically calculated normalized sublattice Zak phase. In all panels, the system parameters are taken as $u=1.2$ and \ch{$v=1.5$}.}
        \label{fig:ssh3wd}
    \end{figure}
\end{center}

In Fig.~\ref{fig:ssh3wd}, the single particle energy spectrum of the SSH3 model is plotted at more general parameter values $u$, $v$, and $w$, along with the corresponding normalized sublattice Zak phase. As predicted, the regime supporting nonzero energy edge modes, i.e., the isolated bands just above and below the middle band in Fig.~\ref{fig:ssh3wd}(a), corresponds precisely to $Z^0_{A,C}=\pi$.

\subsubsection{Extension by applying a nontrivial square-root}

Another method for extending the SSH model is by appropriately defining a new Hamiltonian $H_{\rm sqSSH}$, so that its square yields the real space Hamiltonian of the SSH model. This idea was first envisioned in Ref.~\cite{Ark17}, which considers a general tight-binding Hamiltonian for the square-root model as (in the first-quantized notation)
\begin{equation}
    H_{\rm sqSSH} = \sum_{j=1}^{N} H_j  + \sum_{j=1}^{N-1} \left(T_{j,j+1} +T_{j,j+1}^\dagger \right) ,   
\end{equation}
where $H_j$ and $T_{j,j+1}$ are suitably chosen onsite and nearest-neighbor matrices connecting $j+1$ to $j$ sites, respectively, so that $H_{\rm sqSSH}^2$ produces the target Hamiltonian, e.g., the SSH model. By explicitly squaring $H_{\rm sqSSH}$, we obtain
\begin{eqnarray}
    H_{\rm sqSSH}^2 &=& \sum_{j=1}^N H_j' + \sum_{j=1}^{N-1} \left(T_{j,j+1}' +T_{j,j+1}'^\dagger  \right) \nonumber \\
    && + \sum_{j=1}^{N-2} \left(\tilde{T}_{j,j+2}' +\tilde{T}_{j,j+2}'^\dagger  \right) , 
\end{eqnarray}
where 
\begin{eqnarray}
    H_j' &=& H_j^2 +T_{j,j+1} T_{j,j+1}^\dagger + T_{j-1,j}^\dagger T_{j-1,j}  , \nonumber \\
    T_{j,j+1}' &=& H_j T_{j,j+1} + T_{j,j+1} H_{j+1} , \nonumber \\
    \tilde{T}_{j,j+2}' &=& T_{j,j+1} T_{j+1,j+2} .
\end{eqnarray}
In the special case of the SSH model, 
\begin{eqnarray}
    H_j' &=& v |j,A\rangle \langle j,B | +h.c. , \nonumber \\
    T_{j,j+1}' &=& w |j,B\rangle \langle j+1,A | , \nonumber \\
    \tilde{T}_{j,j+2}' &=& 0 ,
\end{eqnarray}
where $|j,S\rangle$ is a single particle state at sublattice $S=A,B$ of site $j$. 

\begin{center}
    \begin{figure*}
        \centering
        \includegraphics[scale=0.63]{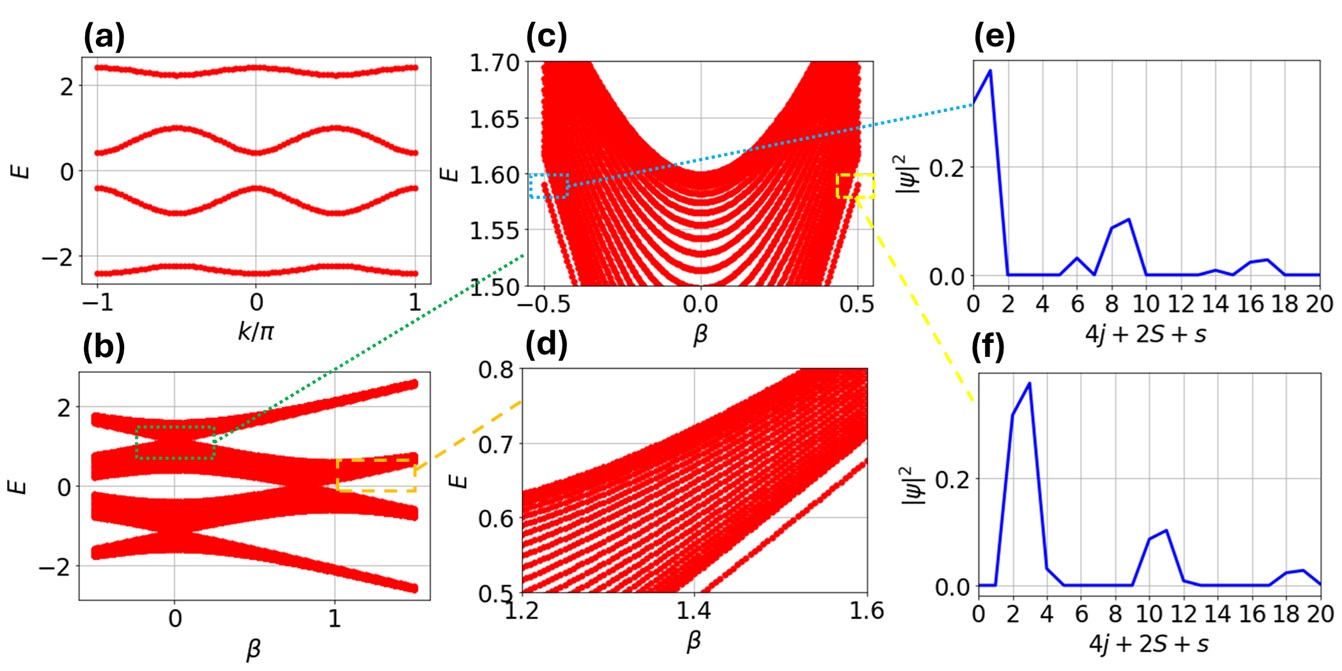}
        \caption{(a) The energy band structure of the square-root SSH model $\mathcal{H}_{\rm sqSSH}(k)$ at $\gamma=\beta=\kappa=1$. (b) The corresponding energy spectrum under OBC at $N=50$, $\gamma=0.6$, $\kappa=1$, and varying $\beta$. Panels (c) and (d) show the zoomed-in views of panel (b) around some relevant bands that support topological edge modes. (e,f) The spatial profile of the edge modes observed in panel (c) at (e) $\beta=-0.5$ and (f) $\beta=0.5$, focusing only on the left-localized edge modes for clarity.}
        \label{fig:sqssh}
    \end{figure*}
\end{center}

The condition $\tilde{T}_{j,j+2}'=0$ implies that $H_{sqSSH}$ necessarily has at least four sites per unit cell, i.e., $H_{sqSSH}^2$ then represents two decoupled copies of the SSH model. The remaining matrices can be chosen as
\begin{eqnarray}
    H_j &=& \sum_{s=\pm 1} (-1)^s \beta \left(|j,A,s\rangle \langle j,A,s | + |j,B,s\rangle \langle j,B,s | \right) \nonumber \\
    && + \sum_{s=\pm 1} (-1)^s \kappa \left(|j,A,s\rangle \langle j,B,s | + |j,B,s\rangle \langle j,A,s | \right) , \nonumber \\
    T_{j,j+1} &=& \sum_{s=\pm 1} (-1)^s \gamma \left( |j,B,-s\rangle \langle j+1,A,s | \right. \nonumber \\
    && \left. - |j,A,s\rangle \langle j+1,B,-s |\right) ,
\end{eqnarray}
where $s=\pm 1$ represent two additional pseudospin/sublattice degrees of freedom. The corresponding momentum space Hamiltonian then reads
\begin{equation}
    \mathcal{H}_{\rm sqSSH}(k) = \left( \begin{array}{cccc}
        \beta & \kappa & 0 & -\gamma e^{-\mathrm{i} k}  \\
        \kappa & \beta & \gamma e^{\mathrm{i} k} & 0 \\
        0 & \gamma e^{-\mathrm{i} k} & -\beta & -\kappa \\
        -\gamma e^{\mathrm{i} k} & 0 & -\kappa & -\beta \\
    \end{array} \right) .
\end{equation}
The energy bands of which can be analytically found as \cite{Ark17}
\begin{equation}
    E_{\mu,\eta} = \eta \sqrt{\beta^2 +\kappa^2 +\gamma^2 + 2\mu \kappa \sqrt{\beta^2 + \gamma^2 \cos^2(k)} } ,
\end{equation}
where $\mu,\eta=\pm 1$ (see Fig.~\ref{fig:sqssh}(a)). By comparing $E_{\mu,\eta}$ with the energy bands of the SSH model, i.e. $E_\pm = \pm \sqrt{v^2+w^2 +2vw\cos(k)}$, as well as by using $\cos^2(k) = \frac{1+\cos(2k)}{2}$, it follows that $\mathcal{H}_{\rm sqSSH}(k)^2$ indeed yields two decoupled copies of the SSH Hamiltonians, which are shifted in energy by $\beta^2 +\kappa^2 +\gamma^2$, with parameter identifications 
\begin{eqnarray}
    \gamma^2 = 4 v w &,& \beta^2 = (v-w)^2, 
\end{eqnarray}
and with the effective Brillouin zone of $[-\pi/2,\pi/2)$.

The system supports two types of band touching points. First, it involves the gap closing of the outer bands, which occurs at $\beta=0$. The second type of gap closing involves the two inner bands, which occurs at $\kappa^2 = \beta^2+\gamma^2$. In Ref.~\cite{Ark17}, two indices encapsulating these two types of band touching points are defined as
\begin{eqnarray}
    \xi = \mathrm{sgn}(\beta^2+\gamma^2-\kappa^2) &,& \tilde{\xi} = \mathrm{sgn}(\beta) .
\end{eqnarray}
The presence of the two types of band touching points is numerically confirmed in Fig.~\ref{fig:sqssh}(b). Indeed, it is clearly observed that, at $\beta=0$, the two upper bands and the two lower bands close their finite energy gaps. Meanwhile, at $\beta=\sqrt{\kappa^2-\gamma^2}=0.8$ for the parameter values under consideration, the gap closes at zero energy, i.e., between the two middle bands.

Similarly to the regular SSH model, $\mathcal{H}_{\rm sqSSH}$ exhibits chiral, particle-hole, and time-reversal symmetries with respect to the operators 
\begin{eqnarray}
    \mathcal{C} = \left(\begin{array}{cccc}
        0 & 0 & 1 & 0  \\
        0 & 0 & 0 & 1 \\
        1 & 0 & 0 & 0 \\
        0 & 1 & 0 & 0 \\
    \end{array}\right) &,& \mathcal{P} = \mathcal{K} \left(\begin{array}{cccc}
        1 & 0 & 0 & 0  \\
        0 & 1 & 0 & 0 \\
        0 & 0 & -1 & 0 \\
        0 & 0 & 0 & -1 \\
    \end{array}\right) , \nonumber \\
    \mathcal{T} &=& \mathcal{C}\mathcal{P} ,
\end{eqnarray}
which also places the system in the BDI class, characterized by an integer invariant. The Zak phase again serves as a suitable invariant for this purpose, which can be analytically calculated as \cite{Ark17}
\begin{equation}
    \gamma_{\mu,\eta} = \begin{cases}
        \left(\eta \tilde{\xi} - \xi\right)\pi/2 & \text{ for }\mu=-1 \\
        \left(\eta \tilde{\xi} - 1\right)\pi/2 & \text{ for }\mu=1 \\
    \end{cases} .
\end{equation}
That there are two distinct expressions for the Zak phase, depending on which band is considered, reflects the fact that there are two independent band gaps, one around zero energy and the other around some finite energy. 
In this case, the presence of edge states inside a particular gap is related to the sum of the Zak phases of the two neighboring bands, i.e.,
\begin{eqnarray}
 \nu_0&=&\frac{\gamma_{-1,1}+\gamma_{-1,-1}}{\pi}=-\xi , \nonumber \\  
 \nu_{\pm} &=& \frac{\gamma_{1,\pm 1}+\gamma_{-1,\pm 1}}{\pi} = \pm \tilde{\xi} -\frac{(1+\xi)}{2} .
\end{eqnarray}
for the zero energy and the finite energy gaps respectively.  

In Fig.~\ref{fig:sqssh}(c) and (d), the zoomed-in views of the energy spectrum shown in Fig.~\ref{fig:sqssh}(b) are presented to highlight the presence of nonzero energy edge modes. In particular, Fig.~\ref{fig:sqssh}(c) shows the zoomed-in view of the top band (corresponding to $\mu=\eta=1$) near $\beta=0$. There, $\tilde{\xi}=1$ and $\tilde{\xi}=-1$ at $\beta>0$ and $\beta<0$ respectively, while $\xi=-1$ in both cases. This regime then corresponds to either $\nu_+=-1$ or $\nu_-=-1$, reflecting a topologically nontrivial regime that supports edge modes, as indeed confirmed in Fig.~\ref{fig:sqssh}(c). In Fig.~\ref{fig:sqssh}(d), the zoomed-in view of the upper middle band (corresponding to $-\mu=\eta=1$) near $\beta =\sqrt{\kappa^2-\gamma^2}$ is presented. Specifically, it focuses on the regime $\beta >\sqrt{\kappa^2-\gamma^2}$, which corresponds to $\xi=1$. In this case, $\nu_0=-1$, which signifies the presence of edge modes inside the energy gap around zero energy, as confirmed in Fig.~\ref{fig:sqssh}(d). While not shown in the figure, the regime $\beta <\sqrt{\kappa^2-\gamma^2}$, which corresponds to the trivial $\nu_0=1$ does not support such edge modes. 

Finally, it is worth noting that while the edge modes are present inside the finite energy gap both at $\beta<0$ and $\beta>0$, these regimes, which are separated by the gap closing at $\beta=0$, are not exactly identical. In particular, Fig.~\ref{fig:sqssh}(e) and (f) present the spatial profile $|\psi|^2(4j+2S+s)$ of the left-localized edge modes observed in Fig.~\ref{fig:sqssh}(c) at $\beta=-0.5$ and $\beta=0.5$ respectively, where $j$ represents the site number, $S=0$ ($S=1$) for pseudospin $+1$ ($-1$), and $s=0$ ($s=1$) for sublattice A (B). There, the two edge modes clearly exhibit qualitatively different profiles, i.e., the largest peak corresponds to the $+1$ ($-1$) pseudospin for the left-localized edge mode at $\beta=-0.5$ and $\beta=0.5$ respectively. This is also consistent with the fact that both edge modes are characterized by distinct topological invariants, i.e., $(\nu_+,\nu_-)=(-1,1)$ and $(\nu_+,\nu_-)=(1,-1)$ at $\beta=-0.5$ and $\beta=0.5$ respectively.   

\subsection{Extending the SSH model from its momentum space Hamiltonian}
\label{SSH3M}
 
\begin{center}
    \begin{figure}
        \centering
        \includegraphics[scale=0.5]{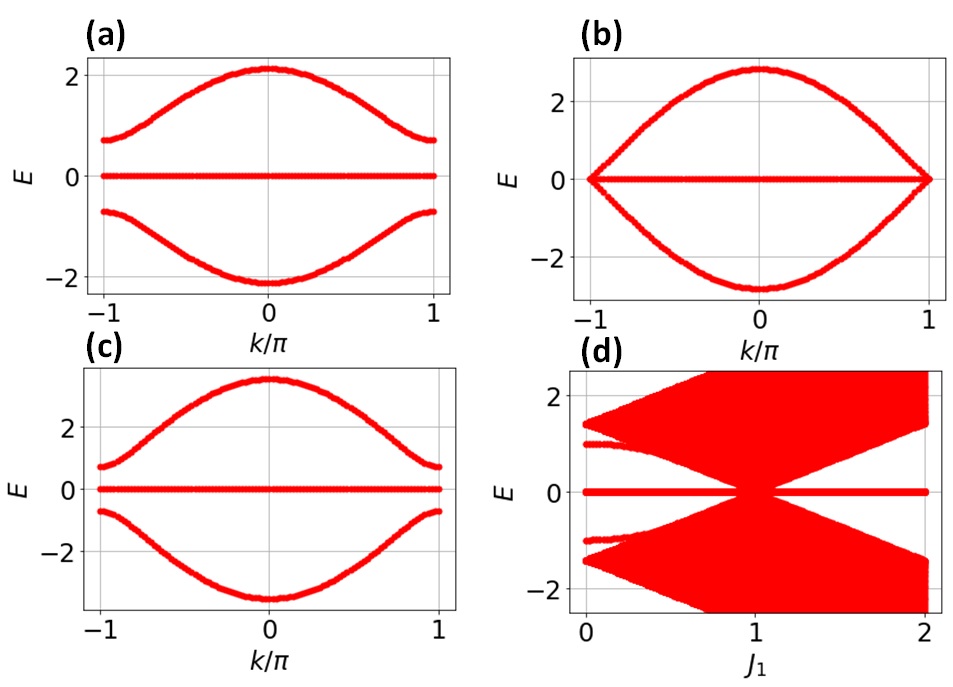}
        \caption{The single particle energy spectrum of the SSH3m model under (a,b,c) PBC and (d) OBC at varying values of $J_1$. The system parameters are chosen as $J_2=1$ (all panels) and (a) $J_1=0.5$, (b) $J_1=1$, (c) $J_1=1.5$. In (d), a system size of $N=100$ is chosen.}
        \label{fig:ssh3mspec}
    \end{figure}
\end{center}

Enlarging the unit cell of the SSH model can also be achieved by directly replacing the $2\times 2$ Pauli matrices that appear in its momentum space Hamiltonian, i.e., Eq.~(\ref{eq:SSHm}), by their higher-level counterparts. This approach was first considered in Ref.~\cite{Ghun24}, which specifically focuses on the trimer extension of the SSH model (to be referred to as the SSH3m hereafter). This amounts to replacing $\sigma_x$ and $\sigma_y$ in Eq.~(\ref{eq:SSHm}) by, respectively,
\begin{eqnarray}
    S_x = \left( \begin{array}{ccc}
        0 & 1 & 0 \\
        1 & 0 & 1 \\
        0 & 1 & 0 \\
    \end{array}\right) &,& S_y = \left( \begin{array}{ccc}
        0 & -\mathrm{i} & 0 \\
        \mathrm{i} & 0 & -\mathrm{i} \\
        0 & \mathrm{i} & 0 \\
    \end{array}\right) ,
\end{eqnarray}
i.e.,
\begin{equation}
    \mathcal{H}_{SSH3m}(k) = \left(J_1+J_2 \cos(k) \right) S_x + J_2 \sin(k) S_y , \label{eq:SSH3m}
\end{equation}
where $J_1$ and $J_2$ denote the intra- and inter-cell hopping amplitudes respectively. In contrast to Eq.~(\ref{ssh3modmom}), the three-band momentum space Hamiltonian arising from the above procedure preserves the chiral symmetry, described by the operator $\mathcal{C} = \mathrm{diag}(1,-1,1)$. The presence of the chiral symmetry is also evident from the typical band structure shown in Fig.~\ref{fig:ssh3mspec}, which is symmetrical about $E=0$. In addition to the chiral symmetry, the system also respects time-reversal, particle-hole, and inversion symmetries given by $\mathcal{T}=\mathcal{K}$, $\mathcal{P}=\mathcal{K} \mathcal{C}$, and $\mathcal{I}=\left( \begin{array}{ccc}
        0 & 0 & 1 \\
        0 & 1 & 0 \\
        1 & 0 & 0 \\
    \end{array}\right)$ respectively. This places the system in the BDI class, similarly to the regular SSH model. 
    
    Another similarity between the SSH3m and the regular SSH model is the closing of the band gaps at zero energy at equal hopping amplitudes, i.e., $J_1=J_2$ in the SSH3m model as shown in Fig.~\ref{fig:ssh3mspec}(b). However, unlike the regular SSH model, the three-band nature of the SSH3m model implies that one bulk band is necessarily unpaired by the chiral symmetry and must be pinned at zero energy at all values of $J_1$ and $J_2$, i.e., Fig.~\ref{fig:ssh3mspec}(a,c). As a result, SSH3m is unable to support zero energy edge modes. Its topology instead manifests itself as pairs of nonzero energy edge modes that are symmetrically placed about $E=0$ (see the regime $J_1<J_2$ of Fig.~\ref{fig:ssh3mspec}(d)).

Due to the presence of the chiral symmetry, the Zak phase is constrained to a quantized value in the SSH3m model. However, we find that the Zak phase only takes a trivial value, i.e., $0$ or $2\pi$, for all values of $J_1$ and $J_2$ and all bands. This result is unsurprising as the Zak phase is associated with the presence of zero energy edge modes, which cannot exist in the SSH3m model due to the persistence of the zero energy bulk band. \ch{Instead, the topology of the SSH3m manifests itself in the presence of nonzero energy edge modes that lie within the chiral symmetric finite energy gaps.} To correctly predict the nonzero energy edge modes in the SSH3m model, we may again employ the normalized sublattice Zak phase of Eq.~(\ref{eq:nszak}). As the SSH3m model exhibits a simpler momentum space Hamiltonian as compared with the SSH3 model, its normalized sublattice Zak phase can be analytically computed as follows.

First, we note that Eq.~(\ref{eq:SSH3m}) can be analytically diagonalized, which yields
\begin{eqnarray}
    E_0 = 0 &,& E_\pm = \pm \sqrt{2\left(J_1^2+J_2^2+2J_1J_2 \cos(k)\right)} , \nonumber \\
\end{eqnarray}
whilst the (unnormalized) corresponding eigenvectors take the form 
\begin{eqnarray}
    |u_0\rangle = \left(a(k),0,-a^*(k)\right)^T \;,\; |u_\pm\rangle = \left(a(k),\pm 1,a^*(k)\right)^T , \nonumber \\
\end{eqnarray}
where 
\begin{equation}
 a(k)=\frac{J_1+J_2 e^{-\mathrm{i} k}}{\sqrt{2\left(J_1^2+J_2^2+2J_1 J_2\cos(k)\right)}} .   
\end{equation}
Note that unlike Ref.~\cite{Ghun24}, we have chosen a gauge where the second component of the eigenstates (corresponding to sublattice B) is real. Under this choice, the vanishing of the regular Zak phase is more transparent due to the symmetry between the first and the third components of the eigenstates. Moreover, the sublattice-A-projected eigenstate is the same for all three bands and reads $|\tilde{u}_\lambda(k)\rangle =\left(\sqrt{2} a(k),0,0\right)^T$ for all $\lambda=0,\pm$. The normalized sublattice Zak phase can then be written as
\begin{equation}
    Z_{A,B}^\lambda = \frac{\mathrm{i}}{2} \oint \frac{1}{J_1+z} dz ,
\end{equation}
where \ch{we have taken $J_2 e^{-\mathrm{i} k} \rightarrow z$, so that} the integral is over a circular contour of radius $J_2$ in the complex plane, centered around the origin. This is precisely the same type of integral as the Zak phase for the regular SSH model, i.e., Eq.~(\ref{eq:ZakSSH}), which yields 
\begin{equation}
    Z_{A,B}^\lambda = \begin{cases}
        \pi & \text{ for }J_2>J_1 \\
        0 & \text{ for }J_1>J_2 \\
    \end{cases} .
\end{equation}
The value $Z_{A,B}^\lambda=\pi$ (\ch{$Z_{A,B}^\lambda=0$}) corresponds to the topologically nontrivial (trivial) regime that is characterized by the presence (absence) of nonzero energy edge modes. This indeed agrees with the energy spectrum of Fig.~\ref{fig:ssh3mspec}(d), where the edge modes emerge inside the gaps between the zero energy band and the outer energy bands in the regime $J_1<J_2$.  

\section{Extending the SSH model by incorporating additional physical effects}
\label{SSHeff}

One central theme in studies of topological phases is to explore the interplay between topology and various physical effects such as periodic driving, non-Hermiticity, interaction, and nonlinearity. The motivation for such studies are twofold. First, realistic platforms necessarily involve some of these physical effects. Second, these physical effects often give rise to novel topological phenomena. The SSH model serves as an excellent platform for such studies due to its intuitive and analytically tractable structure. 

\ch{While the various physical effects above are equally interesting and relevant in studies of topological phases, their analysis incorporates significantly different strategies. Indeed, treating periodically driven systems typically involves studies of the (unitary) time-evolution operator rather than the system Hamiltonian. Meanwhile, the fact that non-Hermitian systems generally support complex energies implies that their analysis necessarily takes into account the topology of both energy eigenstates and eigenvalues. Interacting and nonlinear systems are even more subtle to handle. In particular, while the analysis of both periodically driven systems and non-Hermitian systems can be made from some spectral (eigenvalues and eigenstates) structure, this is generally no longer feasible in the interacting and nonlinear setting. Specifically, the exponential increase in the Hilbert space dimension of a typical interacting system implies that obtaining its full energy eigenvalues and eigenstates is extremely challenging, even numerically, let alone utilizing them to define and compute the appropriate topological invariants. Meanwhile, as nonlinear systems are governed by a nonlinear Schr\"{o}dinger equation, linear algebra machinery such as matrix diagonalization and superposition principle are no longer applicable. As a result, studies of interacting and nonlinear systems often employ alternative techniques beyond merely spectral analysis.}

\ch{As this review article focuses on topological characterization based on spectral analysis by building on the intuition of the regular SSH model, we shall only extensively elaborate periodically driven SSH model and non-Hermitian SSH model at the technical level in the following discussion. Other physical modifications of the SSH model that involve interaction effect or nonlinearity will still be briefly discussed for completeness, so as to credit existing efforts in these nontrivial areas and highlight novel physics generated by these effects. However, we shall refrain from going deeper into the technical analysis of these models, as doing so necessitates the introduction of various new terminologies not used before that potentially break the intended flow of this review article. Interested readers are instead advised to directly refer to the cited references or a dedicated review article on interacting systems \cite{Rachel18} and nonlinear systems \cite{Wu03}.}

%Of all physical effects that can be considered, time-periodicity and non-Hermiticity are two of the most extensively studied ones in the context of topological phases, which have led to active research directions of their own termed Floquet topological phases and non-Hermitian topological phases respectively. For this reason, two extensions of the SSH model, i.e., the time-periodic SSH model and non-Hermitian SSH model, will be elaborated quite extensively in the following. Meanwhile, other extensions of the SSH model based on the incorporation of other physical effects will only be briefly discussed.

\subsection{Periodically driven SSH model}

By simply allowing the hopping amplitudes $v$ and $w$ to vary with time, the resulting SSH Hamiltonian becomes time-dependent. Of particular interest is a scenario in which $v(t)$ and $w(t)$ are time periodic, i.e., $v(t+T)=v(t)$ and $w(t+T)=w(t)$ for some period $T$. Such a time periodic system can be studied within the framework of Floquet theory \cite{Shirley65,Sambe73}, which amounts to analyzing the one-period time evolution operator (usually referred to as the Floquet operator) $U_T$ instead of the otherwise time-independent Hamiltonian $H(t)$. Specifically, the two quantities are related via
\begin{equation}
    U_T = \mathcal{T} e^{-\mathrm{i} \int_0^T H(t) dt } ,
\end{equation}
where $\mathcal{T}$ is the time-ordering operator. As $U_T$ is a unitary operator, its eigenvalues take the form $e^{-\mathrm{i}\varepsilon T}$, where $\varepsilon \in \mathbb{R}$ is termed quasienergy due to its close analogue to energy in a typical static system. The corresponding eigenstate is then termed quasienergy eigenstate or Floquet eigenstate. 

To enable analytical treatment, a simple time-periodicity of $H(t)$ is often considered, such as that which takes the form
\begin{equation}
    H(t)=\begin{cases}
        H_1 & \text{ for } \left(t<T/2\right)\text{ mod }T \\
        H_2 & \text{ for } \left(t>T/2\right)\text{ mod }T
    \end{cases}
\end{equation}
for some constant Hamiltonians $H_1$ and $H_2$. In the case of the SSH model, a possible choice for $H_1$ and $H_2$ could be
\begin{eqnarray}
    H_1 = \sum_{j=1}^N v c_{j,B}^\dagger c_{j,A} +h.c. &,& H_2 = \sum_{j=1}^{N-1} w c_{j+1,A}^\dagger c_{j,B}+h.c. , \label{eq:SSHtan} \nonumber \\
\end{eqnarray}
which was considered, e.g., in Ref.~\cite{Tan20}. Within the single particle subspace, the corresponding Floquet operator can then be written as
\begin{eqnarray}
    U_T &=& e^{-\mathrm{i} \left(\sum_{j=1}^{N-1} \tilde{w} |j+1,A\rangle \langle j,B | +h.c. \right)} e^{-\mathrm{i} \left(\sum_{j=1}^N \tilde{v} |j,B\rangle \langle j,A | +h.c. \right)} ,\nonumber \\
\end{eqnarray}
where $\tilde{v}\equiv \frac{vT}{2}$, $\tilde{w}\equiv \frac{wT}{2}$, and $|j,S\rangle$ is a single particle state that represents a particle occupying the sublattice $S=A,B$ of site $j$. Consider now the following three sets of parameter values:

\subsubsection{$\tilde{v}=0$ and $\tilde{w}\neq 0$:} 

It is easily verified that $U_T |1,A\rangle = |1,A \rangle$ and $U_T |N,B\rangle = |N,B \rangle$. That is, the system supports a pair of zero quasienergy edge modes that are exactly given by $|1,A\rangle$ and $|N,B\rangle$.

\subsubsection{$\tilde{v}=\pi$ and $\tilde{w}\neq 0$:} 

By utilizing the Euler-like formula 
\begin{eqnarray}
    e^{-\mathrm{i} \left( \theta |j,B\rangle \langle j,A | +h.c.\right)} &=& \mathcal{I} -\mathrm{i} \sin(\theta) \left(|j,B\rangle \langle j,A | +h.c.\right)  \nonumber \\
    &-& (1-\cos(\theta)) \left(|j,B\rangle \langle j,B | \right. \nonumber \\
    &+& \left.  |j,A\rangle \langle j,A |\right) , \label{eq:eul} 
\end{eqnarray}
where $\mathcal{I}$ is the identity operator, It then follows that $U_T |1,A\rangle = -|1,A \rangle$ and $U_T |N,B\rangle = -|N,B \rangle$. That is, $|1,A\rangle$ and $|N,B\rangle$ now correspond to $\pi/T$ quasienergy eigenstates that are localized near the system's edges. This type of edge mode, which is often referred to as a $\pi$ mode is known to only exist in periodically driven systems, made possible by the physical equivalence between $\pi/T$ and $-\pi/T$ quasienergy values \cite{Jiang11,Bomantara18b,Bomantara20c,Zhu22}. 

\subsubsection{$\tilde{v}=\frac{\pi}{2}$ and $\tilde{w}=\pi$:}

By again using Eq.~(\ref{eq:eul}) and a related formula for $e^{-\mathrm{i} \left( \theta |j+1,A\rangle \langle j,B | +h.c.\right)}$, it can be shown that
\begin{eqnarray}
    U_T \left(|1,A\rangle \pm \mathrm{i} |1,B\rangle\right) &=& \mp \left(|1,A\rangle \pm \mathrm{i} |1,B\rangle\right) , \nonumber \\
    U_T \left(|N,A\rangle \pm \mathrm{i} |N,B\rangle\right) &=& \pm \left(|N,A\rangle \pm \mathrm{i} |N,B\rangle\right) .
\end{eqnarray}
That is, the system supports both zero and $\pi$ edge modes at each edge of the chain.

To uncover the topological nature of such a periodically driven SSH model, it is instructive to analyze the momentum space Hamiltonian $\mathcal{H}(k,t)$. In particular, the time-reversal, particle-hole, and chiral symmetries for a time periodic Hamiltonian read, respectively \cite{Roy17}
\begin{eqnarray}
    \mathcal{T}^{-1} \mathcal{H}(k,t) \mathcal{T} &=& \mathcal{H}^*(-k,T-t), \nonumber \\
    \mathcal{P}^{-1} \mathcal{H}(k,t) \mathcal{P} &=& -\mathcal{H}^*(-k,T-t), \nonumber \\
    \mathcal{C}^{-1} \mathcal{H}(k,t) \mathcal{C} &=& -\mathcal{H}(k,T-t).
\end{eqnarray}
To make these symmetries more transparent, it is convenient to shift the time origin by $t\rightarrow t+T/4$, so that
\begin{equation}
    \mathcal{H}(k,t)=\begin{cases}
        \mathcal{H}_1 & \text{for} \left(|t-T/2|>T/4\right)\text{ mod }T \\
        \mathcal{H}_2 & \text{for} \left(|t-T/2|<T/4\right)\text{ mod }T
    \end{cases} ,\label{eq:fsshm}
\end{equation}
where 
\begin{eqnarray}
    \mathcal{H}_1 &=& v\sigma_x ,\nonumber \\
    \mathcal{H}_2 &=& w\cos(k) \sigma_x +w \sin(k) \sigma_y .
\end{eqnarray}
It is now straightforward to verify that Eq.~(\ref{eq:fsshm}) respects all the above symmetries under the same $\mathcal{T}$, $\mathcal{P}$, and $\mathcal{C}$ as in its static counterpart. Therefore, like its static counterpart, the number of its topological edge modes is characterized by an integer topological invariant. However, since there are two species of topological edge modes, i.e., the zero and $\pi$ modes, a pair of topological invariants are necessary to reveal its full topology.

To define the appropriate topological invariants, we follow the procedure outlined in Ref.~\cite{Asb14}. First, we write the momentum space Floquet operator in the form
\begin{equation}
    \mathcal{U}_T(k) = F G ,
\end{equation} 
where 
\begin{eqnarray}
    F &=& e^{-\mathrm{i} \left(\frac{\tilde{w}}{2} \cos(k) \sigma_x + \frac{\tilde{w}}{2} \sin(k) \sigma_y \right)} e^{-\mathrm{i} \frac{\tilde{v}}{2} \sigma_x} , \nonumber \\
    G &=& \mathcal{C}^{-1} F^\dagger \mathcal{C} .
\end{eqnarray}
In matrix form,
\begin{equation}
    F = \left( \begin{array}{cc}
        A & B \\
        C & D
    \end{array} \right) .
\end{equation}
Finally, we define 
\begin{eqnarray}
    \nu_0 = \frac{1}{2\pi \mathrm{i}} \oint B^{-1} dB &,& \nu_\pi = \frac{1}{2\pi \mathrm{i}} \oint D^{-1} dD ,
\end{eqnarray}
which respectively count the number of pairs of zero and $\pi$ edge modes \cite{Asb14}. 

\begin{center}
    \begin{figure}
        \centering
        \includegraphics[scale=1]{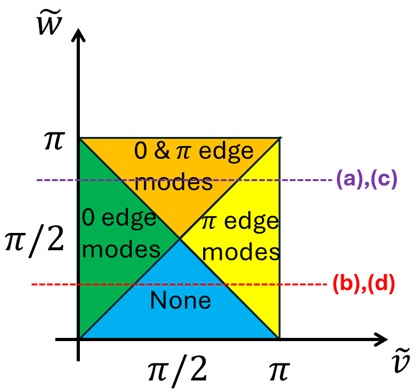}
        \caption{The ``phase diagram" of Eq.~(\ref{eq:SSHtan}) based on the existence of zero and/or $\pi$ edge modes. \ch{The two horizontal lines mark the two specific sets of parameter values analyzed in Fig.~\ref{fig:flosshspec}.}}
        \label{fig:flosshpd}
    \end{figure}
\end{center}

By using the Euler's formula for Pauli matrices, i.e.,
\begin{equation}
    e^{-\mathrm{i} \theta \sigma\cdot \hat{n}} = \cos(\theta) -\mathrm{i} \sin(\theta) \sigma\cdot \hat{n} ,
\end{equation}
we can explicitly write
\begin{eqnarray}
    B &=& -\mathrm{i} \left[ \cos\left(\frac{\tilde{v}}{2}\right)\sin\left(\frac{\tilde{w}}{2}\right) e^{-\mathrm{i} k} - \sin\left(\frac{\tilde{v}}{2}\right)\cos\left(\frac{\tilde{w}}{2} \right) \right] ,\nonumber \\
    D &=& \cos\left(\frac{\tilde{v}}{2}\right)\cos\left(\frac{\tilde{w}}{2}\right) - \sin\left(\frac{\tilde{v}}{2}\right)\sin\left(\frac{\tilde{w}}{2}\right) e^{\mathrm{i} k}
\end{eqnarray}
It then follows that $\nu_0$ and $\nu_\pi$ can be written in the familiar form
\begin{eqnarray}
    \nu_0 &=& \frac{1}{2\pi \mathrm{i}} \oint_{\mathcal{C}_0} \frac{1}{\sin\left(\frac{\tilde{v}}{2}\right)\cos\left(\frac{\tilde{w}}{2} \right)-z} dz , \nonumber \\ 
    \nu_\pi &=& \frac{1}{2\pi \mathrm{i}} \oint_{\mathcal{C}_\pi} \frac{1}{\cos\left(\frac{\tilde{v}}{2}\right)\cos\left(\frac{\tilde{w}}{2} \right)-z} dz ,
\end{eqnarray}
where $\mathcal{C}_0$ and $\mathcal{C}_\pi$ are respectively circular contours in the complex plane of radii $\cos\left(\frac{\tilde{v}}{2}\right)\sin\left(\frac{\tilde{w}}{2}\right)$ and $\cos\left(\frac{\tilde{v}}{2}\right)\cos\left(\frac{\tilde{w}}{2}\right)$, centered around the origin. This implies that
\begin{eqnarray}
    \nu_0 &=& \begin{cases}
        1 & \text{ for }\tan\left(\frac{\tilde{w}}{2}\right)>\tan\left(\frac{\tilde{v}}{2}\right) \\
        0 & \text{ for }\tan\left(\frac{\tilde{v}}{2}\right)>\tan\left(\frac{\tilde{w}}{2}\right) \\
    \end{cases} , \nonumber \\
    \nu_\pi &=& \begin{cases}
        1 & \text{ for }\tan\left(\frac{\tilde{w}}{2}\right)>\cot\left(\frac{\tilde{v}}{2}\right) \\
        0 & \text{ for }\cot\left(\frac{\tilde{v}}{2}\right)>\tan\left(\frac{\tilde{w}}{2}\right) \\
    \end{cases} . \label{eq:wdflossh}
\end{eqnarray}
In particular, as the conditions $\tan\left(\frac{\tilde{w}}{2}\right)>\tan\left(\frac{\tilde{v}}{2}\right)$ and $\tan\left(\frac{\tilde{w}}{2}\right)>\cot\left(\frac{\tilde{v}}{2}\right)$ are not mutually exclusive, there exists a regime in which zero and $\pi$ edge modes coexist. Within $(v,w)\in[0,\pi)\times [0,\pi)$, Eq.~(\ref{eq:wdflossh}) translates to the phase diagram of Fig.~\ref{fig:flosshpd}.

\begin{center}
    \begin{figure}
        \centering
        \includegraphics[scale=0.38]{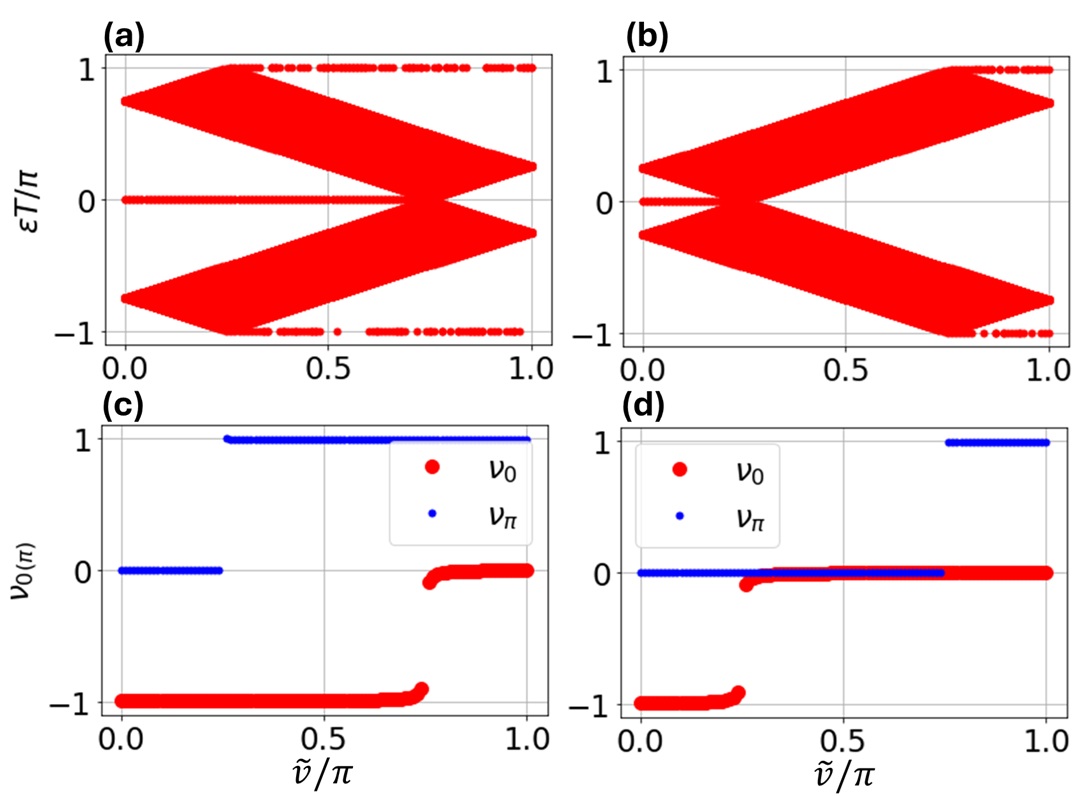}
        \caption{\ch{(a,b) The quasienergy spectra of the periodically driven SSH model at varying $\tilde{v}$ along, respectively, the purple ($\tilde{w}=3\pi/4$) and red ($\tilde{w}=\pi/4$) lines of Fig.~\ref{fig:flosshpd} ($N=100$ in both cases). (c,d) The corresponding winding numbers $\nu_0$ and $\nu_\pi$ at varying $\tilde{v}$.} }
        \label{fig:flosshspec}
    \end{figure}
\end{center}

\ch{For completeness, Fig.~\ref{fig:flosshspec} shows the numerically calculated quasienergy energy spectra of the periodically driven SSH model, along with the corresponding winding numbers $\nu_0$ and $\nu_\pi$ for two different sets of parameter values. In particular, Fig.~\ref{fig:flosshspec}(a,c) covers three regimes, one of which supports only zero edge modes, another supports both zero and $\pi$ edge modes, and the other supports only $\pi$ edge modes. This corresponds to sweeping the parameter $\tilde{v}$ along the purple line in Fig.~\ref{fig:flosshpd}. Meanwhile, Fig.~\ref{fig:flosshspec}(b,d) sweeps the parameter $\tilde{v}$ along the red line in Fig.~\ref{fig:flosshpd}, covering a regime that supports only zero edge modes, a regime that does not support edge modes, and a regime that supports only $\pi$ modes. In all cases, the numerically calculated $\nu_0$ and $\nu_\pi$ correctly predict the presence/absence of zero edge modes and $\pi$ edge modes respectively.}

\subsection{Non-Hermitian SSH model}

It is well-known that a Hamiltonian describing a closed quantum system is Hermitian in nature. This restriction underlies various defining properties of quantum states, such as the realness of all energy eigenvalues, the completeness of the corresponding energy eigenstates, and the unitary nature (reversibility) of the time evolution operator. By allowing a system's Hamiltonian to be non-Hermitian, which may arise, e.g., in classical systems \cite{Gao21,Helb20,Wang23}, as an approximate description of some open quantum system \cite{Rocc22}, or as an effective Hamiltonian that governs the dynamics of some bosonic operators \cite{Mc18,Bomantara25,Bomantara25b}, some of these properties may no longer hold. This leads to the modification of the spectral theory for analyzing the system's Hamiltonian, which in turn affects its topological classification. As a result, topological effects that have no Hermitian counterparts may often arise in non-Hermitian systems.    

One of the simplest non-Hermitian variants of the SSH model is described by
the Hamiltonian (using first-quantized notation)
\begin{eqnarray}
    H_{\rm NH} &=& \sum_{j=1}^{N} \left[ \left(v + \frac{\gamma}{2} \right) |j,B\rangle \langle j,A | + \left(v - \frac{\gamma}{2} \right) |j,A\rangle \langle j,B | \right] \nonumber \\ 
    && + \left[ \sum_{j=1}^{N-1} w |j+1,A\rangle \langle j,B | +h.c. \right] , \label{eq:nhSSHr}
\end{eqnarray}
where $\gamma$ reflects some nonreciprocity in the intracell hopping amplitude, i.e., inequality between the A to B and B to A sublattice hopping amplitudes. The model Eq.~(\ref{eq:nhSSHr}) has been proposed and extensively studied in Ref.~\cite{Yao18}, which further takes into account an additional longer-range hopping term. For simplicity, we do not consider such a term in this section, i.e., we set $t_3=0$ in Eq.~(1) of Ref.~\cite{Yao18}, as Eq.~(\ref{eq:nhSSHr}) suffices to demonstrate the role of non-Hermiticity in enriching the topology of the SSH model. Without loss of generality, we further assume that all parameter values $v$, $w$, and $\gamma$ are positive-valued.

By applying the transformation Eq.~(\ref{eq:rtmtrans}), the corresponding momentum space Hamiltonian reads
\begin{equation}
    \mathcal{H}_{\rm NH}(k)= (v+w\cos(k)) \sigma_x +(w\sin(k) +\mathrm{i} \frac{\gamma}{2}) \sigma_y . \label{eq:nhSSH}
\end{equation}
the eigenvalues of Eq.~(\ref{eq:nhSSH}) can be readily obtained as
\begin{equation}
    E_\pm (k) = \pm \sqrt{v^2 +w^2 -\frac{\gamma^2}{4} +w \left(2v \cos(k)+ \mathrm{i} \gamma \sin(k)\right) },  
\end{equation}
which are generally complex. It is then instructive to define the energy winding number (with respect to a reference complex energy $E$) as \cite{Okuma20}
\begin{eqnarray}
    \mathcal{W}_{E} &=& \frac{1}{2\pi \mathrm{i}} \oint \left[ (E_+-E)^{-1} dE_+ + (E_- - E)^{-1} dE_- \right] \nonumber \\
    &=& \frac{1}{2\pi \mathrm{i}} \oint \frac{1}{v^2+w^2-\frac{\gamma^2}{4}+ z -E^2} dz ,
\end{eqnarray}
where the integral is over an elliptical contour, centered around the origin in the complex plane, with the maximum real axis coordinate at $2wv$ and the maximum imaginary axis coordinate at $w\gamma$. Explicitly carrying out the integral yields  
\begin{equation}
    \mathcal{W}_E = \begin{cases}
        1 & \text{ for } (v-w)^2<\frac{\gamma^2}{4}+E^2 <(v+w)^2 \\
        0 & \text{otherwise}
    \end{cases} . \label{eq:nhsshwd}
\end{equation}
Note that such a nontrivial winding number at the level of the energy eigenvalues rather than the eigenstates is unique to non-Hermitian systems. In the literature, $\mathcal{W}_E$ characterizes the point-gap topology of the non-Hermitian Hamiltonian. Specifically, the system has a nontrivial point-gap topology if there exists a reference complex energy $E$ such that $\mathcal{W}_E=1$. Otherwise, i.e., if $\mathcal{W}_E=0$ for all values of $E$, the system is topologically trivial. 

\begin{center}
    \begin{figure}
        \centering
        \includegraphics[scale=0.7]{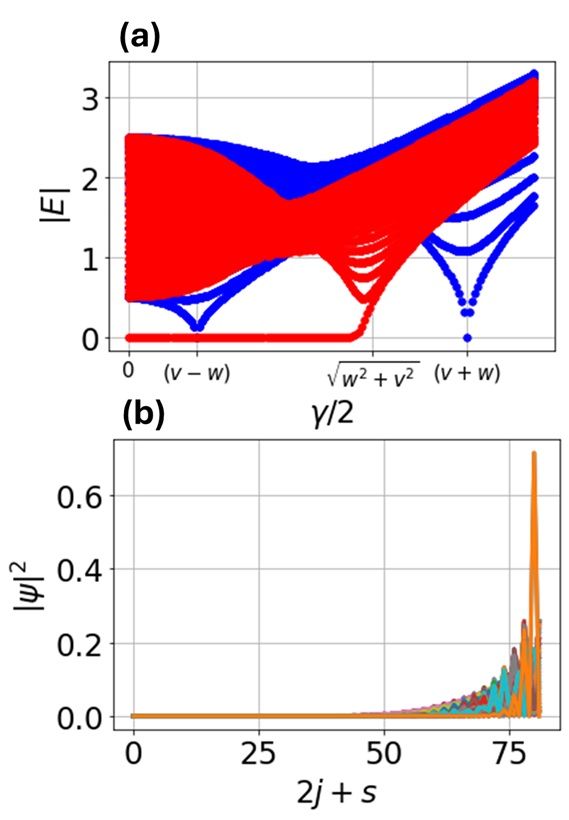}
        \caption{\ch{(a) Side-by-side comparison of the energy spectra of the non-Hermitian SSH model under OBC (red) and PBC (blue). (b) The spatial profile of all corresponding eigenstates at a fixed $\gamma=0.4$. In all panels, the other system parameters are taken as $N=41$, $v=1$, and $w=1.5$.} }
        \label{fig:nhsshspec}
    \end{figure}
\end{center}

Physically, the point-gap topology determines if a non-Hermitian system exhibits the non-Hermitian skin effect \cite{Okuma20}, characterized by the breakdown of the usual bulk--boundary correspondence and the localization of all bulk states near a system's boundary \cite{Kunst18,Yao18b,Lee16,Alvarez18,Xiong18,Lin23}. For the non-Hermitian SSH model considered above, Eq.~(\ref{eq:nhsshwd}) shows that the system is always in a nontrivial point-gap topological phase regardless of the system parameters $v$, $w$, and $\gamma$. Indeed, given a specific set of parameter values for $v$, $w$, and $\gamma$ such that $(v+w)^2-\frac{\gamma^2}{4}>0$, one may choose a real-valued $E$ that lies between $\sqrt{{\rm min}\left[0,(v-w)^2-\frac{\gamma^2}{4}\right]}$ and $\sqrt{(v+w)^2-\frac{\gamma^2}{4}}$ to yield $\mathcal{W}_E=1$. If these parameters instead satisfy $(v+w)^2-\frac{\gamma^2}{4}<0$, one may instead choose an imaginary-valued $E$, the norm of which lies between $\sqrt{{\rm min}\left[0,\frac{\gamma^2}{4}-(v-w)^2\right]}$ and $\sqrt{\frac{\gamma^2}{4}-(v+w)^2}$ to achieve it. Therefore, the non-Hermitian SSH model of Eq.~(\ref{eq:nhSSHr}) exhibits the non-Hermitian skin effect for all values of $v$, $w$, and $\gamma$. \ch{This can also be directly verified by numerically plotting the corresponding energy spectra under OBC and PBC side-by-side. Indeed, as shown in Fig.~\ref{fig:nhsshspec}(a), clear discrepancy between the PBC spectrum (blue marks) and OBC spectrum (red marks) is observed. In particular, the PBC spectrum yields a gap closing at two points $\gamma/2 = v\pm w$, whilst the OBC spectrum only supports one gap closing point at $\gamma/2=\sqrt{w^2+v^2}$, which actually corresponds to a topological phase transition as further elaborated below. }

In the original proposal of Ref.~\cite{Yao18}, the localization of all eigenstates of the non-Hermitian SSH model is uncovered through another means. First, for $2v>\gamma$, a similarity transformation is applied to Eq.~(\ref{eq:nhSSHr}) according to
\begin{equation}
    \overline{H}_{\rm NH} = S^{-1} H_{\rm NH} S ,
\end{equation}
where 
\begin{eqnarray}
    S= \sum_{j=1}^N \left(r^{j-1} |j,A\rangle \langle j,A | + r^{j} |j,B\rangle \langle j,B | \right) &,& r=\sqrt{\frac{2v+\gamma}{2v-\gamma}} . \nonumber \\
\end{eqnarray}
The resulting Hamiltonian becomes
\begin{eqnarray}
    \overline{H}_{\rm NH} &=& \sum_{j=1}^N \overline{v} |j,B\rangle \langle j,A | + \sum_{j=1}^{N-1} w |j+1,A\rangle \langle j,B | +h.c. , \nonumber \\ 
\end{eqnarray}
where $\overline{v} = \sqrt{v^2-\frac{\gamma^2}{4}}$. That is, $\tilde{H}_{\rm NH}$ is simply the regular (Hermitian) SSH model with an effective intracell hopping parameter of $\sqrt{v^2-\frac{\gamma^2}{4}}$. The usual diagonalization procedure can then be carried out to obtain all of its eigenstates, which are delocalized in nature except for a pair of zero energy edge modes that exist in the regime $w>\overline{v}$. The eigenstates of the original non-Hermitian model $H_{\rm NH}$ are related to those of $\overline{H}_{\rm NH}$ through multiplication by $S$. Specifically, let $|\overline{u}_E\rangle =\sum_{j=1}^N \left(a_j |j,A\rangle \langle j,A | + b_j |j,B\rangle \langle j,B |\right) $ be an eigenstate of $\overline{H}_{\rm NH}$ corresponding to energy $E$. The associated eigenstate of $H_{\rm NH}$ is then obtained as
\begin{equation}
    |u_E\rangle = S|\overline{u}_E\rangle = \sum_{j=1}^N \left(r^{j-1} a_j |j,A\rangle \langle j,A | + r^j b_j |j,B\rangle \langle j,B |\right) .
\end{equation}
Since $r>1$, it follows that $r^{j-1}a_j$ and $r^j b_j$ increase exponentially with the lattice site $j$. Consequently, in a sufficiently long lattice, all eigenstates of $H_{\rm NH}$ is localized near the right end, i.e., non-Hermitian skin effect is established. \ch{The right-edge localization of all eigenstates can also be verified numerically by plotting the spatial profile (recall Eq.~(\ref{eq:sprof}) for its definition) of all eigenstates of the non-Hermitian SSH model at fixed parameter values, as shown in Fig.~\ref{fig:nhsshspec}(b).}

For $2v<\gamma$, a different similarity transformation could be applied with respect to 
\begin{eqnarray}
    \tilde{S}= \sum_{j=1}^N \left(\tilde{r}^{j-1} |j,A\rangle \langle j,A | + \tilde{r}^{j} |j,B\rangle \langle j,B | \right) &,& \tilde{r}=\sqrt{\frac{\gamma+2v}{\gamma-2v}} . \nonumber \\
\end{eqnarray}
which results in
\begin{eqnarray}
    \tilde{H}_{\rm NH} &=& \sum_{j=1}^N \left[ \frac{\overline{\gamma}}{2} |j,B\rangle \langle j,A | - \frac{\overline{\gamma}}{2} |j,A\rangle \langle j,B | \right] \nonumber \\
    && + \sum_{j=1}^{N-1} w |j+1,A\rangle \langle j,B | +h.c. , \nonumber \\ 
\end{eqnarray}
where $\frac{\overline{\gamma}}{2}= \sqrt{\frac{\gamma^2}{4}-v^2}$. Similarly to the case $2v>\gamma$, any eigenstate of $H_{\rm NH}$ is related to that of $\tilde{H}_{\rm NH}$ according to 
\begin{equation}
    |u_E\rangle = \tilde{S} |\tilde{u}_E\rangle = \sum_{j=1}^N \left(\tilde{r}^{j-1} a_j |j,A\rangle \langle j,A | + \tilde{r}^j b_j |j,B\rangle \langle j,B |\right) ,
\end{equation}
which is again localized near the right end. This analysis confirms that non-Hermitian skin effect indeed exists at all parameter values as predicted by the point-gap topological analysis. 

\subsection{Other physical modifications of the SSH model}

As the original SSH model involves only nearest-neighbor hopping terms, adding some longer-range hopping terms serves as a natural extension of the model. This was considered, e.g., in Refs.~\cite{Hsu20,Gonz19,Mal23,Cha25}. In particular, Ref.~\cite{Hsu20} specializes to the case of adding next-nearest-neighbor hopping terms only, whereas Ref.~\cite{Gonz19,Mal23,Cha25} considers general long-range hopping which includes next-nearest-neighbor hopping terms and beyond. It is found that in the presence of next-nearest-neighbor hopping terms, the resulting model can support a regime with a larger winding number of $2$, in addition to that of $0$ and $1$ winding number as typically expected in the regular SSH model \cite{Hsu20}. \ch{In the presence of longer-range hopping terms beyond the next-nearest-neighbor type, such a winding number remains a suitable topological invariant for certain models such as those studied in Refs.~\cite{Mal23,Cha25}, where even larger winding numbers are achievable. For some other models, such as that studied in Ref.~\cite{Gonz19}, the winding number no longer serves as an appropriate topological invariant.}

%Meanwhile, such a winding number may no longer serve as a suitable topological invariant in the presence of longer range hopping terms beyond the next-nearest-neighbor type \cite{Gonz19}.

As real life systems typically involve a large number of interacting particles, the single-particle description of the SSH model may not fully apply. Motivated by this, a number of studies have considered a type of extended SSH model that incorporates an extra interacting term. For example, Ref.~\cite{Liberto16} studies a bosonic version of the SSH model with additional Hubbard interaction, focusing on the presence of only two particles for simplicity. One notable finding of this study is the possibility of edge modes to be present in a parameter regime that is otherwise topologically trivial in the non-interacting limit. In Ref.~\cite{Salvo24}, a fermionic version of the SSH model in the presence of Hubbard-like interaction is explored. Unlike Ref.~\cite{Liberto16}, a many-body system beyond two particles is considered in Ref.~\cite{Salvo24}, but it focuses on its low-energy description to still enable some analytical treatment. The main findings of Ref.~\cite{Salvo24} include the construction of a topological invariant for such an interacting 1D topological system, as well as the physical interpretation of such an invariant.   

Apart from some specific classes of systems such as that considered in Ref.~\cite{Salvo24}, interacting systems are typically very difficult to handle, even numerically. Indeed, while the Hilbert space of a non-interacting system scales linearly with the system size, that of an interacting system scales exponentially instead. Unless there exist symmetries or relevant approximations that can be exploited, numerical studies of interacting systems are limited by the number of particles that is accessible by existing computers, which is typically only of the order of tens. In some cases, nonlinearity may serve as a suitable approximation to interaction effect. 

In view of the above, some recent studies \cite{Tul20,Ezawa21} have also been dedicated to explore the effect of nonlinearity on the SSH model, i.e., by adding a state-dependent term to the Hamiltonian. Despite being effectively non-interacting, solving a nonlinear Schr\"{o}dinger equation is no easy feat, since superposition principle breaks down and the usual linear algebra analysis such as matrix diagonalization procedure cannot be directly carried out. Techniques such as an iterative method \cite{Tul20}, self-consistent analysis \cite{Tul20}, and/or numerical evolution studies of some initial states \cite{Ezawa21} are instead employed to tackle such nonlinear problems. For the specific type of nonlinearity considered in Refs.~\cite{Tul20,Ezawa21}, i.e., onsite nonlinearity, new phenomena that arise in the resulting nonlinear SSH model include the nonlinearity-induced modification of the Zak phase, solitons of topological origin, as well as the presence of incomplete band structure (energy bands that do not span the whole Brillouin zone). 

\section{Concluding remarks}
\label{conc}

The SSH model is a versatile topological model that can be extended in various ways to yield and explore different types of topological phases. Indeed, as demonstrated in Sec.~\ref{TISSH}, some minimal models describing higher-dimensional topological phases such as topological insulators and Weyl semimetals share a similar topological origin as the SSH model in the continuum limit. Moreover, Sec.~\ref{HOSSH} further discussed that a class of higher-order topological phases can be obtained by simply stacking multiple copies of the SSH chains. While keeping its 1D nature intact, several variations of extended SSH models can also be devised by enlarging the size of one unit cell. In particular, this unit cell enlargement can be made through different processes, such as by changing the periodicity of the hopping amplitudes from two to three sites, applying some nontrivial square-rooting procedure, or replacing the Pauli matrices by their higher-level counterparts, all of which have been previously detailed in Sec.~\ref{SSH3} and Sec.~\ref{SSH3M}. Finally, as elaborated in Sec.~\ref{SSHeff}, incorporating physical effects such as periodic driving, non-Hermiticity, long-range hopping, interaction, and nonlinearity gives rise to another class of extended SSH models that yield richer topological phenomena. \ch{In Table~\ref{tab:summary}, the various extended SSH models elaborated above are summarized in terms of the method of extension used, protecting symmetries, and their appropriate topological invariants.}   

\begin{table*}
    \centering
    \begin{tabular}{|c|c|c|c|}
    \hline 
        \textbf{Extension method} & \textbf{Variant} & \textbf{Protecting symmetries} & \textbf{Topological invariant} \\
        \hline 
        \hline 
        Discretizing modified Dirac Hamiltonian  &  QWZ model & None & Chern number ($C_\pm$) \\
        \cline{2-4}
        in higher dimensions & 3D Topological insulator & Time-reversal & $Z_2$ parity ($\nu$) \\
        \cline{2-4}
         & Weyl semimetal & Broken time-reversal or inversion & Chirality ($\chi$) \\
         \hline 
         Coupling multiple SSH models & Second-order & Chiral and spatial symmetries & Winding number ($\mathcal{W}_{xy}$)  \\
         along orthogonal directions & topological insulator & & \\
         \hline
         Modifying the sublattice structure & SSH3 model & Point chiral symmetry & Normalized sublattice \\
         & & & Zak phase ($Z_{A,C}^\lambda$) \\
         \cline{2-4}
         & SSH3m model & Chiral, time-reversal, & Normalized sublattice \\
         & & inversion, and particle-hole & Zak phase \\
         \cline{2-4}
         & Square-root SSH model & Chiral, time-reversal, & Sums of Zak phases \\
         & & and particle-hole & ($\nu_0,\nu_\pm$) \\
         \hline 
         Incorporating physical effects & Periodically driven SSH model & Chiral, time-reversal, & A pair of winding \\
         & & and particle-hole & numbers ($\nu_0$, $\nu_\pi$) \\
         \cline{2-4}
         & Non-Hermitian SSH model & Chiral, time-reversal, & Energy winding number \\
         & & and particle-hole & ($\mathcal{W}_E$) and Zak phase \\
         \hline
    \end{tabular}
    \caption{\ch{A summary of all the extended SSH models discussed in this review article.}}
    \label{tab:summary}
\end{table*}

The various elements above can also work in tandem to yield even richer extended SSH models. In existing studies, some extended SSH models that incorporate multiple of these elements have indeed been uncovered. For example, in Ref.~\cite{Bomantara19}, periodic driving and the procedure of stacking of multiple SSH chains are combined to yield a Floquet second-order topological insulator that yields both zero and $\pi$ corner modes. Meanwhile, Ref.~\cite{Ghun25} studies a periodically driven version of the SSH3m of Eq.~(\ref{eq:SSH3m}). In Ref.~\cite{Bomantara22}, the square-rooting procedure of Ref.~\cite{Ark17} is extended to the time-periodic realm to yield a square-root Floquet topological insulator. Reference~\cite{Zhou22} further utilizes a combination of periodic driving, non-Hermiticity, and the square-rooting procedure of Ref.~\cite{Bomantara22} to yield a family of $q$th-root non-Hermitian Floquet topological insulators. Without involving periodic driving, Ref.~\cite{He21} considers the presence of non-Hermiticity in the SSH3 model of Eq.~(\ref{ssh3mod}), while Ref.~\cite{Liu19} studies the presence of non-Hermiticity in the second-order topological insulating model that arises from stacking multiple SSH chains.

Apart from its use as a starting model in theoretical studies, the SSH model and its extensions have also been subjects to experimental explorations in various platforms. Such experiments typically do not simply realize the SSH model or its extensions, instead they mainly seek after exotic physical observables in such models when implemented in different platforms such as photonics \cite{Cheng19,Iva23,On24,Klauck21,Roberts22,Aravena22,Upa24,Henri20}, acoustic waveguides \cite{Chen22,Coutant21,Li23}, mechanical systems \cite{Mir24,Thatcher22}, ultracold atoms \cite{Ata13,Xie19}, or superconducting circuits \cite{Deng22,Yous22,Cai19}. 

\ch{The aforementioned experimental platforms utilize different strategies for implementing the target model (the SSH model or its variation). For instance, in photonic and acoustic waveguides, the Schr\"{o}dinger equation on a discrete quantum lattice is simulated by the classical wave equation describing an array of waveguides under appropriate conditions. In such a waveguide lattice, multiple species of nearest-neighbor hopping amplitudes can be achieved by controlling the distance between two adjacent waveguides. Since the propagation direction serves as the effective time dimension, a time-periodic Hamiltonian can be easily simulated through appropriately engineering the shape of each waveguide. This in turn enables the experimental observation of $\pi$ edge modes \cite{Cheng19}, i.e., which are unique to the periodically driven SSH model with no static counterparts. Another advantage of the waveguide setup is the fact that it is inherently classical, allowing for physical effects such as nonlinearity and non-Hermiticity to be easily implemented. In particular, onsite nonlinearity could be induced through Kerr effect, whilst non-Hermiticity arises through gain/loss mechanism. The latter was utilized in Ref.~\cite{Iva23} to realize a pair of coupled non-Hermitian SSH models, which was further shown to exhibit phenomena resembling the quantum Zeno effect.}

\ch{Other experimental platforms such as ultracold atoms and superconducting circuits offer means to realize the SSH model and its extensions as fully quantum systems. Such platforms are particularly advantageous for observing physical quantities that are related to topological invariants, as well as for utilizing topological edge modes for quantum technological applications. For example, Ref.~\cite{Ata13} was able to directly measure the Zak phase of the SSH model in an optical lattice of ultracold atoms via Bloch oscillations and Ramsey interferometry. Meanwhile, Ref.~\cite{Xie19} utilized mean chiral displacement measurement to extract the topological invariant associated with the SSH4 model, i.e., a further extension of the SSH3 model that involves four distinct nearest-neighbor hopping amplitudes. In Ref.~\cite{Deng22}, various topological models, including the SSH model, are realized in a chain of superconducting qubits (gmons), which are coupled through the use of resonators and inductive couplers. By adiabatically tuning the latter, a topological edge mode initially localized in one end of the lattice can be transferred to the other end.}

Moving forward, the SSH model is expected to remain a fundamental building block of numerous topological systems for both theoretical and experimental studies. Indeed, as the number of possible topological phases of matter in the universe is virtually endless, the various extended SSH models proposed in existing studies cannot exhaust all of them. In fact, some of these extended SSH models may serve as new starting points for developing even more complex topological phases and/or for exploring their potential technological applications. For example, starting from the periodically driven version of the SSH model, Ref.~\cite{Tan20} further devised a scheme to utilize its edge modes as qubits and transfer them from one place to another. Meanwhile, Ref.~\cite{Cheng22} further modifies the periodic driving scheme to yield a model that supports an even more exotic type of edge mode, then explicitly realizes it experimentally in a system of acoustic waveguides. Finally, having highlighted the significance of the SSH model and its extensions for the studies of topological phases in this review article, it is hoped that the various case studies elucidated above further stimulate ideas for future research.

%\begin{acknowledgements}
%This work was supported by the Deanship of Research
%Oversight and Coordination (DROC) at King Fahd University of Petroleum \& Minerals (KFUPM) through project No.~EC221010.
% \end{acknowledgements}

%\appendix

\end{document}